\documentclass[12pt,preprint]{aastex}
\usepackage{graphicx,epsfig}
\usepackage{lscape}
\usepackage{rotating}
\newcommand{\bq}{\begin{equation}}
\newcommand{\eq}{\end{equation}}

\def\gtsim{\lower.5ex\hbox{$\buildrel > \over\sim$}}
\def\ltsim{\lower.5ex\hbox{$\buildrel < \over\sim$}}

\def\apjl{ApJL}
\def\apj{ApJ}

\def\mnras{MNRAS}

\def\aj{AJ}
\def\aap{A\&A}

\def\nat{Nature}
\def\pasp{PASP}
\shorttitle{SN2008am: A Super Luminous IIn SN at z=0.234}
\shortauthors{Chatzopoulos,Wheeler,Robinson,Vinko,Miller,Foley,Perley,Quimby,Yuan,Akerlof,Bloom}
\begin{document}
\title
{SN2008am: A Super-Luminous Type IIn Supernova}
\author{E. Chatzopoulos\altaffilmark{1}, J. Craig Wheeler
\altaffilmark{1}, J. Vinko\altaffilmark{2}, R. Quimby\altaffilmark{3}, 
E. L. Robinson\altaffilmark{1}, A. A. Miller\altaffilmark{4}, R. J. Foley\altaffilmark{5,8},
D. A. Perley\altaffilmark{4}, F. Yuan\altaffilmark{6}, C. Akerlof\altaffilmark{6},
and J. S. Bloom\altaffilmark{4,7}}
\authoremail{manolis@astro.as.utexas.edu}
\altaffiltext{1}{Department of Astronomy, University of Texas at Austin, Austin, TX, USA. E-mail:
manolis@astro.as.utexas.edu}
\altaffiltext{2}{Department of Optics and Quantum Electronics, University of Szeged, Szeged, Hungary}
\altaffiltext{3}{Division of Physics, Mathematics and Astronomy, California Institute of Technology, Pasadena, 
CA 91125, USA}
\altaffiltext{4}{Department of Astronomy, University of California, Berkeley, CA, 94720-3411, USA}
\altaffiltext{5}{Harvard-Smithsonian Center for Astrophysics, 60 Garden Street, Cambridge, MA, 02138, USA}
\altaffiltext{6}{University of Michigan, Randall Laboratory of Physics, 450 Church St., Ann Arbor, MI, 48109-1040, USA}
\altaffiltext{7}{Sloan Research Fellow}
\altaffiltext{8}{Clay Fellow}

\begin{abstract}

We present observations and interpretation of the Type IIn supernova SN~2008am 
discovered by the ROTSE Supernova Verification Project (RSVP). SN~2008am peaked at 
approximately -22.3~mag at a redshift of $z =$~0.2338, giving it a peak luminosity of
$\sim 3 \times 10^{44}$~erg~s$^{-1}$ and making it one of the most luminous supernovae 
ever observed. The total radiated energy is $\simeq 2 \times 10^{51}$~erg. The host 
galaxy appears to be an SB1 of normal luminosity ($M_{r^{\prime}} \sim$~-20) with
metallicity $Z \sim $~0.4~$Z_{\odot}$.
ROTSE upper limits and detections constrain the rise time to be $\sim$~34 days in the 
rest frame, significantly shorter than similar events, SN~2006gy and SN~2006tf. Photometric 
observations in the ultraviolet, optical and infrared bands ({\it J,H,K$_{s}$}) constrain 
the SED evolution. We obtained six optical spectra of the supernova, five on the 
early decline from maximum light and a sixth nearly a year later plus a very late-time 
spectrum ($\sim$~2 yr) of the host galaxy. The spectra show no evidence for broad 
supernova photospheric features in either absorption or emission at any phase. 
The spectra of SN~2008am show strong Balmer-line and He I $\lambda$5876 \AA\ emission with intermediate 
widths ($\sim$ 25 \AA) in the first $\sim$~40 days after optical maximum. The width 
formally corresponds to a velocity of $\sim$~1000~km~s$^{-1}$. We examine a variety of models 
for the line wings and conclude that multiple scattering is most likely, implying 
that our spectra contain no specific information on the bulk flow velocity. We examine 
a variety of models for the ROTSE light curve subject to the rise time 
and the nature of the spectra, including radioactive decay, shocks in optically-thick 
and optically-thin circumstellar media (CSM) and a magnetar. The most successful model is
one for which the CSM is optically-thick and in which diffusion of forward shock-deposited
luminosity gives rise to the observed light curve.
The model suggests strong mass loss and a greater contribution from the interaction of the forward shock 
with optically thick CSM than from the reverse shock. Diffusion of the shock-deposited energy from the forward shock
is found to be important to account for the rising part of the light curve.
Although there are differences in detail, SN~2008am appears to be closely related to
other super-luminous Type IIn supernovae, SN~2006gy, SN~2006tf and perhaps SN~2008iy,
that may represent the deaths of very massive LBV-type progenitors and for which the
luminosity is powered by the interaction of the ejecta with a dense circumstellar medium. 

\end{abstract}


\vskip 0.57 in

\section{INTRODUCTION}\label{intro}

The Texas Supernova Search (TSS; Quimby et al. 2005) and its successor, the ROTSE-Supernova Verification Project (RSVP;
Yuan et al. 2007), discovered a new class of super-luminous supernovae (SLSNe). The advantage of the TSS/RSVP project is
that it is essentially free of selection bias and the limits of a targeted search. The automated wide field (3.4 square degree)
ROTSE-III telescopes (Akerlof et al. 2003) scan the whole sky, looking for transients down to $\sim$19~mag. They do not focus
on pre-selected galaxies nor omit galaxy centers. The first TSS/RSVP discoveries in this new class of SLSNe were
SN~2005ap (Quimby et al. 2007a), SN~2006gy (Quimby 2006; Smith et al. 2007), SN~2006tf (Quimby, Castro \& Mondol 2007; Quimby et al. 2007b;
Smith et al. 2008) and SN~2008es (Yuan et al. 2008b; Gezari et al. 2009; Miller et al. 2009). 
These exceptionally luminous supernovae are rare, with an estimated rate of
$\sim 2.6 \times 10^{-7}$~events~Mpc$^{-3}$~yr$^{-1}$ (Quimby et al. 2009). The SLSNe introduced new modes of
stellar death. Traditional ideas about the mechanisms that can power supernova luminosity were found to be inadequate to explain
the observed properties of these events. 

The small, but growing, sample of SLSNe is heterogeneous. Some show strong
emission lines of hydrogen in their spectra close to maximum light (SN~2006gy, SN~2006tf, SN~2008fz, SN~2008iy) and typically
belong to the Type~IIn subclass; some show hydrogen in later phases and a linear decline of the light curve expressed in magnitudes (SN~2008es). 
Others may show no hydrogen at all (SN~2005ap, SCP06F6).
For the super-luminous Type~IIn events, the energy generation mechanism is very likely the interaction between the ejecta and 
a circumstellar medium (CSM) that was shed by the progenitor star in
the years prior to the explosion (Chevalier \& Fransson 2003).
SN~2006gy triggered discussions about the possibility of nearby pair-instability
supernovae (Smith et al. 2007). Such models proved unsatisfactory for SN~2006gy 
and many other SLSNe, but may account for SN~2007bi (Gal-Yam et al. 2009; Young et al. 2010). 
Even for some events that do not show clear signs of CSM interaction, simple radioactive decay 
diffusion models (Arnett 1982; Valenti et al. 2008) have proven inconsistent with the observations 
(Quimby et al. 2007a; Gezari et al. 2009; Quimby et al. 2009). Other mechanisms that can account for the large
luminosity have been proposed: interaction between expelled shells (Woosley, Blinnikov \& Heger 2007); interaction between a 
GRB-like jet and the progenitor envelope (Young et al. 2005; Gezari et al. 2009); a buried magnetar (Kasen \& Bildsten 2010; Woosley 2010);
or a very energetic core-collapse explosion (Umeda \& Nomoto 2008; Moriya et al. 2010). In addition, the possibility that many Type~IIn SNe
(of normal or high luminosity) have been spectroscopically confused with radio-quiet low-luminosity blazars has been discussed 
(Filippenko 1989). All recent SLSNe candidates have shown spectroscopic features that are more consistent with SNe.

In the present work, we report on SN~2008am discovered by RSVP (Yuan et al. 2008a). The paper is organized as 
follows. In \S 2 we present the photometric and spectroscopic observations of SN~2008am and discuss the evolution of its spectral 
energy distribution (SED). In \S 3 we consider the nature of the emission-line features and models to account for the
line profiles, and in \S 4 we discuss the applicability of 
various models to account for the light curve. Finally, in \S 5 we summarize our conclusions.

\section{OBSERVATIONS}\label{obs}

\subsection{{\it The host of SN2008am}}\label{host}

The host of SN~2008am is SDSS J122836.32+153449.6 that appears to be a faint, extended object in the Sloan Digital
Sky Survey, with an 
r$^{\prime}$ magnitude of $\sim$20 (see Figure 1). The position of the 
centroid of the host is $\alpha =$~12$^{h}$28$^{m}$36.3$^{s}$ and $\delta =$~+15$^{d}$34$^{m}$50$^{s}$ and 
its redshift is $z =$~0.2338 (Yuan et al. 2008a; \S2.4). 
We estimate the luminosity distance of the SN to be 1130 Mpc assuming a $\Lambda$-CDM cosmology with
$\Omega_{\Lambda} =$~0.73, $\Omega_{M} =$~0.27 and $H_{0} =$~73~km~s$^{-1}$~Mpc$^{-1}$.
On that basis, the host has an absolute magnitude $M_{r^{\prime}} \sim$~-20~mag and thus is not a dwarf galaxy.
The morphology of this galaxy is unknown, but the shape of its SDSS
photometric SED and the optical spectrum
showing narrow emission features of H, [NII], [OII] and [OIII] 
agree well with an SB1 galaxy template spectrum (Figure 9; \S 2.4).
The metallicity of the host, estimated from the flux ratios of these narrow emission features, is sub-solar 
($Z \sim $~0.4~$Z_{\odot}$), similar to the hosts of other SLSNe (Neill et al. 2010; Stoll et al. 2010; \S 2.4).

\subsection{{\it Imaging and photometry}}\label{phot}

 SN~2008am was discovered in unfiltered ROTSE-IIIb images on 2008 Jan 10.4 UT ($MJD =$~54475.4; Yuan et al. 2008a)
when it had an unfiltered magnitude of 18.7. The position of the SN in the ROTSE images was determined to be
$\alpha =$~12$^{h}$28$^{m}$36.25$^{s}$ and $\delta =$~+15$^{d}$34$^{m}$49.1$^{s}$, slightly offset 
from the center of its host that appears in the SDSS catalog (Figure 1). 
ROTSE-IIIb continued to gather unfiltered
photometric data for $\sim$200 days after discovery. The ROTSE-IIIb photometry is summarized in Table 1. 
The ROTSE data are shown in Figure 2 along with the detection limits over the course of the photometry. The ROTSE-IIIb photometry includes 
an early upper limit and data points during the rising phase of the supernova light
curve that can be used to constrain the explosion date of the SN. To determine the explosion date, we converted the ROTSE magnitudes to
luminosities assuming zero bolometric correction, and fit radioactive decay diffusion models (\S 4.1) to the resulting light curve.
The fitting parameters are the effective diffusion time, the
nickel mass, the explosion date and a parameter that controls the gamma-ray leakage. The best
fit radioactive-decay diffusion model is shown in Figure 3.
This fitting process yields an explosion date of $MJD_{expl} =$~54438.8$\pm$~1,
approximately 18 days prior to the first real detection in the observed frame and about 14 days in the rest frame.
We note that this model was employed only to determine the explosion date of the SN and may not account for the real
physical situation in SN~2008am (see \S 4.1).
The ROTSE-IIIb maximum
occurred at $MJD =$~54480.4 (2008 Jan 15.0), about 34 rest-frame days after the explosion.

At the distance of 1130 Mpc for the ROTSE unfiltered peak magnitude of 18.0~mag, the absolute ROTSE peak magnitude of SN~2008am is
$-$22.3~mag (uncorrected for extinction).
The ROTSE response curve peaks in the red and it is calibrated to the USNO-B1.0 system (Smith et al. 2003). There is always 
a slight offset from
a true R magnitude due to the fact that the shape of a supernova SED is very different from the reference stars used by ROTSE, but, to a good
approximation, this absolute peak magnitude is close to the real absolute R magnitude of the event. This establishes
that SN~2008am is a super-luminous event; one of the most luminous supernovae ever discovered,
placing the supernova in the family of SLSNe together with SN~1992ar (Clocchiatti et al. 2000), SN~1999as (Hatano et al. 2001),
SN~2003ma (Rest et al. 2009), SN~2005ap, SN~2006gy, SN~2006tf, SN~2008es, SN~2008fz
(Drake et al. 2010), SN~2008iy (Miller et al. 2010),
SN~2007bi (Gal-Yam et al. 2009), SCP06F6 (Barbary et al. 2009) and recently-discovered luminous 
PanSTARRS transients (SN~2009kf, Botticella et al. 2010; SN~2010gx, Pastorello et al. 2010).

 SN~2008am was followed up with photometric observations ranging from the ultraviolet (UV) through the infrared (IR). The Peters
Automated Infrared Imaging Telescope (PAIRITEL; Bloom et al. 2006) obtained {\it J}, {\it H} and {\it K$_{s}$} band photometry over a 25
day period (Table 2). The PAIRITEL {\it J}, {\it H} and {\it K$_{s}$} fluxes are calibrated to the Two Micron All
Sky Survey (2MASS) catalogue (Skrutskie et al. 2006).
 The infrared light curves of SN~2008am over this phase appear to be flat and are probably heavily contaminated by the host.
In the {\it K$_{s}$}-band, especially, the detections are most probably indicative of the host rather than the SN. Thus we refer to them only
as upper limits throughout this work. Although the infrared observations were obtained for only a small portion of the life of the SN,
they can be used to better constrain the SED towards the infrared region for some phases. Contribution by dust IR radiation
to the observed {\it J}, {\it H} and {\it K$_{s}$} fluxes cannot be ruled out, but we have made no allowance for that process.
 
 The Palomar 60-inch (P60) telescope obtained optical photometry in the bands g, r, i$^{\prime}$, and 
z$^{\prime}$ for a period of $\sim$330 days in the rest frame. 
Photometry on the P60 frames was performed using the aperture photometry 
routines in {\it IRAF\footnote[1]{IRAF is distributed by the National Optical Astronomy 
Observatory, which is operated by the Association of Universities for Research in Astronomy (AURA) 
under cooperative agreement with the National Science Foundation.}}. 
The aperture radius was set to be 20 pixels (7.57~arcsec), and the 
background level was measured in an annulus with 30 pixels (11.36~arcsec) inner and 50 pixels (18.93~arcsec)
outer radii, centered on the source. The P60 data were calibrated via five local tertiary standard stars having
Sloan g$^{\prime}$, r$^{\prime}$, i$^{\prime}$ and z$^{\prime}$ magnitudes in the SDSS catalogue (Table~3) . 
The applied P60 filters were $g$ and $r$ 
(similar to the Thuan-Gunn filters; Thuan \& Gunn, 1976), and i$^{\prime}$ and z$^{\prime}$ that resemble the SDSS filters 
(Fukugita et al. 1996), although systematic differences exist between the P60 filters and their standard
counterparts (Cenko et al. 2006). Due to the lack of standard fields observed simultaneously with the SN field, 
only approximate calibration of the P60 photometry was possible. As a first approximation, the SN magnitudes were 
tied to the g$^{\prime}$, r$^{\prime}$, i$^{\prime}$ and z$^{\prime}$ magnitudes of the local tertiary standards assuming 
only a zero-point shift and no color term. Table 4 details the results of the P60 photometry of SN~2008am.
The error caused by the neglect of the color term was investigated by comparing the observed differential magnitudes and 
colors of the local standards with their catalogued magnitudes. Only the data from the $g$ filter are 
affected by the lack of the color term, systematically at a level of $\sim 0.1$ mag.  No significant magnitude shifts were 
detected in any other filters. The resulting g$^{\prime}$, r$^{\prime}$, i$^{\prime}$, z$^{\prime}$ AB-magnitudes of SN 2008am 
were then transformed to fluxes adopting the filter parameters and flux zero points of Cenko et al. (2009). 
The lack of the color term caused less than 1~$\sigma$ error in the g$^{\prime}$ fluxes, where $\sigma$ is the random error of the photometry.
The SN fluxes were subsequently corrected for host contamination and interstellar reddening. The host correction was done
by removing the host fluxes obtained from SDSS. For the interstellar reddening we used the interstellar absorption maps
of Schlegel et al. (1998), giving $E(B-V)_{MW} =$~0.025~mag and the Milky Way reddening law parametrized by Fitzpatrick \& Massa (2007) 
adopting $R_{V} =$~3.1. Throughout this work, we use only the Milky Way reddening correction to obtain the final photometry for SN~2008am. 
The 50 to 180 day slope of the P60 g, r, i$^{\prime}$ lightcurve is estimated to be 0.0065$\pm$0.0006~mag~d$^{-1}$. 
The z$^{\prime}$ light curve is somewhat flatter over the same period. 

{\it Swift} photometry was obtained by the Ultraviolet/Optical Telescope (UVOT; Roming
et al. 2005) covering the first $\sim$~80 rest-frame days after maximum. 
The conversions between Swift magnitudes and fluxes were
computed based on the calibration using the Pickles stellar templates
(Poole et al. 2008) instead of the GRB templates included in the Swift CALDB.
The result of the {\it Swift} photometry is detailed in Table 5.
In the latest three epochs the {\it UVW1}, {\it UVM2} and {\it UVW2} fluxes
seem to level off. This may be due to the increasing contribution from the flux of the host galaxy
relative to the decreasing SN flux at later epochs.

The rest-frame light curves of SN~2008am for all the available photometric bands are shown in Figure 4. 
The top left panel shows the PAIRITEL IR J and H light curves, the top right panel the {\it Swift} UVOT optical and UV light curves
and the bottom panel the ROTSE unfiltered and P60 optical light curves.
The data sets have been offset for clarity, with longer wavelength on top and shorter wavelength toward the bottom of each panel. 
The broad photometric wavelength coverage allows us to better constrain the SED of the 
SN.

SN~2008am was observed with the VLA at 8.46 GHz on 2008 Feb 19.37 UT, approximately 30 rest frame days after
optical maximum (Chandra \& Soderberg 2008). 
No source was detected at the SN position above 120~$\mu$Jy, which can be considered as a 3-$\sigma$ upper limit.
{\it Swift} X-ray Telescope (XRT; Burrows et al. 2005) images 
of SN~2008am were obtained for the six epochs in parallel with the UVOT observations. 
No X-ray source was detected at the position of the SN, after coadding all XRT observations.
Assuming a power-law spectrum with spectral index of $-2$, the X-ray flux upper limit
was estimated to be $\sim 10^{-13}$~erg~cm$^{-2}$~s$^{-1}$ (or, equivalently $\sim 10^{43}$~erg~s$^{-1}$) at $\sim$~50 rest frame
days since explosion.

\subsection{{\it The SED of SN~2008am}}\label{sed}

The availability of multi-band simultaneous photometry for some epochs helps us construct SEDs for SN~2008am and thus
study the evolution and basic properties of the event. Our criteria for selecting the photometric epochs for which we
constructed the SEDs were two: close sampling in time and maximum possible wavelength coverage. 
Those criteria led to the choice of 10 epochs for which we constructed the photometric
SEDs. For four of those epochs we had available UV+Optical+IR observations (hereafter UVOIR), for four only optical (P60 data; 
hereafter OPT) and for two (08-01-30 and 08-02-25)
Optical+IR (see Figure 4). 

For the 08-02-25 epoch we interpolated between
the previous (08-02-16) and the next (08-03-14) epoch for which we had UV data. The error of this interpolated
value was estimated using error propagation analysis for a linear fit.
We note that the UV flux for the interpolated 08-02-25 epoch is uncertain 
since the UV light curve decline of emission line objects like SN~2008am is not well constrained at later times.
Adding the interpolated UV data to the Optical+IR data gave us a total of 5 epochs of UVOIR data and 1 epoch (08-01-30) for OIR data that
we will later use to estimate the pseudo-bolometric light curve (LC) of the event.

Although we see no sign of classic supernova photospheric P Cygni lines (see \S 2.4 and \S 3), the
blue continua of the spectra are consistent with thermal emission. This
emission could arise in shocked CSM that is modestly optically-thick ($\tau \sim$~1) to absorption. To produce a quantitative diagnostic, 
we fit single temperature black-body curves to the set of the five UVOIR and one OIR photometric SEDs in order to determine
effective black-body temperatures in the rest frame. We included the reddening corrections discussed in \S 2.2. 
The black-body curve fits were done in the rest-frame of the SN. Bolometric luminosities were then derived from the integral of the 
flux from the corresponding black body at the adopted distance of the supernova. Effective black-body radii were derived
from the luminosity and black-body temperature. We did not attempt to fit the OPT epochs with single temperature black-body curves 
as the uncertainties of the fit would be large, given the lack of UV data. We estimated lower limits for the bolometric luminosity 
in these four epochs by integrating the observed SED.

Figure 5 shows the photometric SEDs of SN~2008am for the six selected epochs together with the best-fit (lowest $\chi^{2}$) 
black-body curve in each case. Figure 6 shows two examples of UVOIR SEDs and their black-body curve fits and nearly contemporaneous
spectral continua at +22d and +33d after maximum, respectively. Black-body fits to the spectroscopic data were also
done in the rest frame and included the reddening corrections of \S 2.2. The spectra of SN~2008am were scaled to the simultaneous 
photometry in each case, before any reddening corrections were applied. The temperatures derived from the spectral fits agree with
those derived from the fits to the SEDs within the errors at these two epochs (Figure 6).

Table 6 summarizes the characteristics of the black-body 
fits to the UVOIR and OIR SEDs of SN~2008am, the effective black-body temperature, $T_{bb}$, the effective black-body radius, $R_{bb}$, and 
the derived bolometric luminosity, $L_{bol}$. 
Figure 7 presents the evolution of $T_{bb}$ (upper panel), $R_{bb}$ (middle panel), and $L_{bol}$ (lower panel). 
The filled triangles correspond to the fits to the UVOIR and OIR SEDs and the open circles to those of four early HET and
Keck spectra (the fifth, obtained on 2008 Jan 30.3 has a slightly anomalous slope that we attribute to observing conditions; \S 2.4). 
The single temperature black-body fits to the SEDs of SN~2008am show considerable 
uncertainty and scatter. The photometric points scatter around the best-fit black-body curves
and around the spectral continua. Differences between the spectral and photometric results at
similar epochs may be attributed to the fact that the optical spectra do not accurately constrain the maximum of the black-body curve 
and that we assumed a single black-body to fit the SEDs and not more complex models. At the latest epochs there could be other effects 
(for example, dust formation) that affect the final estimates. The NIR data are always in excess with respect to the fitted curves. 
That may be an indication that the single-temperature black-body models do not accurately represent the emission properties of 
SN~2008am, but it is also possible that the NIR data are still somewhat contaminated by the host galaxy and 
only represent upper limits to the corresponding supernova flux. Another possibility 
is that the NIR excess is a sign of early warm dust emission (Meikle et al. 2007; Mattila et al. 2008; Kotak et al. 2009; Fox et al. 2010).
The IR data are thus especially uncertain, but these data have relatively little weight in the black-body fits. The black-body
temperatures and radii derived here are only indicative of the general conditions and their trends, and not to be treated as 
quantitatively precise, nor as evidence that the emission is truly black body.

The effective black-body temperatures are in the range of 10,000-12,000 K, as presented in Table 6 and Figure 7, and are constant in time within 
the scatter.
The pseudo-bolometric light curve implies a total radiated energy of about 10$^{51}$~ergs. Other estimates of the 
total radiated energy will be given in \S 4 based on several physical models.
We will argue in \S 4.4 that the energy powering SN~2008am is most probably the interaction between the ejecta and the 
CSM and thus the estimates of $T_{bb}$, $R_{bb}$ and $L_{bol}$ do not correspond to an expanding photosphere coincident 
with the ejecta of the SN. The ejecta of the SN and the CSM shocks may be hidden behind an optically-thick CSM. In this case the values of $T_{bb}$, 
$R_{bb}$ and $L_{bol}$ refer to conditions in an optically-thick CSM with $\tau \sim$~1 representing the effective photosphere of the CSM.

\subsection{{\it Spectroscopy}}\label{spec}

We acquired a total of seven spectra of SN~2008am and its host galaxy spanning $\sim$~2 years after discovery.
Four of them were taken with the HET Low Resolution Spectrograph (LRS; Hill et al. 1998)
on 2008 Jan 29.3 UT, 2008 Jan 30.3, 2008 Feb. 18.3 and 2008 Feb. 25.3, corresponding to
+11, +12, +27 and +33 rest-frame days after maximum, respectively. 
Two other spectra were obtained with the Keck-I Low Resolution Imaging Spectrograph (LRIS;
Oke et al. 1995) on 2008 Feb. 12.0 UT (+22 days after maximum in rest-frame) and 
2009 Mar. 31.0 (+352 rest-frame days after maximum). 
Another Keck-LRIS spectrum of the host galaxy was taken on 2010 Jan. 09.0 UT, 
+554 rest-frame days after maximum, when the transient had faded below
the detection limit.  
All spectra were reduced in the standard way using {\it IRAF}. The instrumental resolution
was $\sim$17 \AA ~for the HET spectra and $\sim$6 \AA ~for the Keck spectra. 

We will refer to the spectra obtained
within a month after maximum as ``early-phase" and the +352d Keck-spectrum as ``late-phase,"
respectively. All spectra were corrected for
Milky Way reddening (as described in \S 2.2) and scaled to contemporaneous photometric
data. The epochs of the early spectroscopic observations relative to the light curve
are shown in Figure 3. 

Figure 8 shows the spectral evolution of SN~2008am. Note that the first HET spectrum 
(obtained on 2008 Jan 29.3) is plotted for completeness, but omitted from further
analysis because of the availability of the second spectrum (taken one day later,
on Jan 30.3) that was obtained during better observing conditions giving
better S/N ratio. The Jan 30.3 spectrum has a somewhat discrepant slope for reasons we have been unable
to resolve, but the continuum slope of
the spectra does not enter into our analysis except for perhaps a very small effect on
line profiles. The 2008 Jan 29.3 spectrum was used in the black-body fits (\S2.3; Figure 7).

The spectra were deredshifted by $z =$~0.2338 determined from
the narrow emission features in the host spectrum. This value
is very close to the one derived by Yuan et al. (2008a) based on the early-phase
HET spectra.

The early-phase spectra show prominent features of H  (H$\alpha$, H$\beta$
and H$\gamma$). The He~I 5876 \AA ~line is detected in the Feb 12 Keck spectrum and
with a smaller S/N ratio in the Feb 25 HET spectrum. Na D may also contribute to this feature, 
but we were unable to make a definite identification. 

The HET data had inferior seeing and lower spatial 
resolution of the spectrograph that made it impossible to fully resolve and separate the SN and the host. 
Given those instrumental differences between HET-LRS and Keck-LRIS we chose to analyze the spectra uncorrected
for host extinction so that we could treat all the data in a uniform way.
The host probably contributes to some of the detected emission lines; [OII] 3727 \AA ~and 
[OIII] 5007 \AA ~are undoubtedly present, and [OIII] 4959 \AA ~can also be weakly detected
in all HET spectra. Fortunately, the host contribution to the
continuum should be much less, because the +554d Keck spectrum, which is attributed
to the host galaxy, shows mainly a flat, low continuum. 

Due to better resolution and seeing, the host subtraction was possible 
for the early-phase and late-phase Keck spectra. The small visible spatial extension 
of the galaxy required a background region defined as
close to the SN as possible to achieve good background subtraction. Most of the galactic
forbidden emission lines have been removed successfully; however, this process 
resulted in a slight oversubtraction in the core of H$\alpha$ for the early-phase
Keck spectrum, forming an ``absorption" dip on top of the line. 
That feature is due to the reduction process and should not be
interpreted physically (see below). Although we show the host-subtracted early Keck
spectrum for comparison, we note again that we do not use it for our analysis since host subtraction
cannot be performed for the HET spectra given the different spatial resolution.

The late-phase Keck spectrum is also dominated by a broadened H$\alpha$ line,
but all other SN features have already faded below detectability. 
The shape of H$\alpha$ strongly suggests that the transient was still
visible at +352 days after maximum. 
A few narrow features at H$\beta$ and around $\sim$5000 \AA~ also appear that
are probably due to contribution from the host. 

The host spectrum, obtained with Keck-LRIS at +554 rest-frame days after maximum (Figure 9),
shows the usual narrow emission features of galaxies with ongoing star-formation.
Beside Balmer lines, we identified 
[OII] 3727 \AA,~ [OIII] 4959, 5007 \AA,~ [NII] 6548, 6583 \AA ~and He I 5876 \AA.
There are also indications for the [SII] 6716, 6731 \AA ~features, but that region
is heavily contaminated by tellurics, preventing a definite identification. 
All lines as well as the shape of the continuum (Figure 9; \S2.1) can be very well matched by an
SB1 galaxy template (Kinney et al. 1996). 
The metallicity of the host was estimated
by computing the line flux ratios and the spectral indices N2 and O3N2 defined by Pettini \& Pagel (2004).
These resulted in an oxygen abundance of $12 + \log(O/H)$ = $8.38 \pm 0.15$, significantly
below the solar abundance value ($\sim 8.7 \pm 0.1$). This oxygen abundance suggests a
sub-solar metallicity for the host, about $Z ~\sim ~0.4 ~Z_\odot$. There is growing
evidence that SLSNe, such as SN~2008am, occur mostly in metal-deficient hosts
(Neill et al. 2010; Stoll et al. 2010).

None of the SN~2008am spectra show any sign of broad features characteristic of
most SNe during the photospheric phase. 
There is no sign of P Cygni profiles. 
The Balmer lines in the early-phase spectra have FWHM 
$\sim 25$ \AA ~in the rest frame, which are usually referred to as ``intermediate-width" lines
(e.g. Stathakis \& Sadler 1991). Such intermediate-width
features are common characteristics of Type IIn SNe (e.g. Schlegel 1990).
Based on these observed features, 
SN~2008am is certainly a member of the Type IIn subclass. The early-phase spectra
are similar to those of SN~1988Z (Stathakis \& Sadler 1991), SN~1995G (Pastorello
et al. 2002) and the early spectra of SN~2006gy (Smith et al. 2010), all having blue continua with 
intermediate-width Balmer lines and without strong P Cygni profiles. 
In Figure 10 we plot 
the +22d Keck spectrum together with the spectrum of SN~1998Z at +39d after maximum
light that had intermediate-width lines with no P Cygni components
(Stathakis \& Sadler 1991, Turatto et al. 1993, Aretxaga et al. 1999). 
A spectrum of SN~2006gy taken 19 rest-frame days after maximum with the HET by one of
us (R.Q.) is also shown in Figure 10 for comparison. Although some differences in
the fine details exist, we argue below that SN~2008am and SN~2006gy 
exhibited remarkable similarities both in their spectral appearance and 
evolution. 

Figure 11 shows the evolution of the emission lines that can be attributed to 
SN~2008am. For the early Keck spectrum we give both the total flux profile (dashed red curve) 
and the line after host subtraction (solid red curve).
The left panel presents H$\alpha$, the middle panel H$\beta$
and the right panel the He~I/Na~D feature that is only weakly detected.
Dotted vertical lines mark the rest wavelength of the features.
In Table~7 we list the basic parameters derived for H$\alpha$ and H$\beta$;
the shift of the line center with respect to its rest-frame position
($\Delta \lambda_0$, expressed in km s$^{-1}$), 
the full width at half maximum (FWHM), integrated flux (F) and 
equivalent width (EW) based on fitting of Lorentzian profiles (more line
profile models will be examined below). 

In the earliest spectra, the line centers of both H$\alpha$ and H$\beta$ 
are redshifted with respect to their rest-frame position (determined from the
narrow emission features of the host, as mentioned above). 
The average redshift is 
115 $\pm$ 40 km s$^{-1}$ for both Balmer lines, 
consistent with some multiple-scattering models (Chugai 2001; \S3).
On the other hand, in the late-phase Keck spectrum 
the cores of the Balmer lines are slightly blueshifted by $-60$ km s$^{-1}$.
The H$\alpha$ profile in this spectrum looks narrower close to the peak, suggesting the
presence of an unresolved narrow component, similar to the late-time
H$\alpha$ of SN~2006gy (Smith et al. 2010). This slight blueshift with
respect to the rest-frame position may be explained by rotation of the
host, but might also arise from the effects of multiple scattering (\S3).
  
The early-phase spectra have rest-frame FWHM line widths of 
$\sim$~22 \AA ~and $\sim$~23 \AA ~for H$\alpha$
and H$\beta$, respectively. 
Formally expressing these widths in terms of velocity (see \S 3 for a discussion
on the effects of scattering) the FWHM are $\sim$1000 km s$^{-1}$ and 1400 km s$^{-1}$
for H$\alpha$ and H$\beta$ respectively.
H$\beta$ seems
to be broader in terms of velocity than H$\alpha$. After removing the host flux, H$\alpha$ appears
broader in the early Keck spectrum than in the uncorrected spectrum or in any HET spectra,
but a more detailed analysis showed that this is an
artifact due to the depressed amplitude of the line core as a result
of the host galaxy removal. 

The HET spectra contain fluxes from both the SN and the host,
thus the EWs and line fluxes are higher than those derived from
the host-subtracted early-phase Keck-spectrum.
Comparing the numbers in Table~7 and correcting
for the small host oversubtraction in the Keck spectrum, 
the host contribution to EWs and line fluxes in the HET spectra 
are estimated to be $\sim$17 \AA ~and 
$\sim 5 \times 10^{-16}$ erg s$^{-1}$ cm$^{-2}$, respectively. 

The evolution of the line strengths and EWs for SN~2008am is very similar to that
presented by Smith et al. (2010) for SN~2006gy and to
several other Type IIn SNe. Shortly after maximum light the EW of H$\alpha$
is $\sim$30 - 40 \AA, rising to $\sim$300
\AA ~at late phases. The H$\alpha$ integrated line fluxes, on the 
other hand, tend to decline in time. SNe are quite heterogeneous
regarding the evolution of this parameter according to 
Smith et al. (2010), but the majority show a similar decline to that of
SN~2008am. The line fluxes of H$\beta$ also show a declining
trend toward later phases, but the EW of H$\beta$ is roughly constant with
time.

During the early phase, the F(H$\alpha$)/F(H$\beta$) ratio 
is $\sim 2.3 \pm 0.6$, which is close to the expected 
value in case B recombination (Osterbrock, 1989). 
SN~2006gy showed a very similar flux ratio close to maximum light
(Smith et al. 2010). This ratio increases up to $\sim17$ 
during the late phases in SN~2008am. This is also in accord with
the observations of other SNe. An
even higher flux ratio ($\sim 30$) was observed for SN~2006gy (Smith et
al. 2010) and SN~2006tf (Smith et al. 2008) as well as for 
the strongly interacting Type IIn 
SN~1988Z (Turatto et al. 1993). 

The He~I 5876 \AA ~line clearly shows up in the early-phase Keck
spectrum and is also weakly detected in the HET spectra, although
the latter data are noisy. The rest-frame FWHM of this line is formally measured to
be 1800 $\pm$ 600 km s$^{-1}$. This line-width is typical for
He~I lines in other Type IIn events (e.g. SN~2005la, Pastorello
et al. 2008). The total integrated flux in the early Keck spectrum
is measured to be $\sim$~5.5 $\times$ 10$^{-16}$ 
erg~s$^{-1}$~cm$^{-2}$ by fitting a Lorentzian profile. It is possible
that Na~D ($\lambda \lambda$ 5890, 5896 \AA) absorption contaminates this feature making it weaker. 
The F(H$\alpha$)/F(He~I 5876) flux ratio is $\sim 3$, probably
reflecting the temperature/ionization conditions in the line-forming region.  
We do not detect the He~II 4686 \AA ~line in any of our spectra.    

\section{THE NATURE OF THE EMISSION LINES IN SN~2008am}\label{lines}

We conclude from the lack of broad SN features in the early-phase
spectra of SN~2008am that the observed spectra are probably
formed by CSM interaction, in which the SN ejecta collide with 
a dense CSM cloud surrounding the progenitor. In \S4 we consider CSM interaction
and other mechanisms to produce the light curve. Presuming CSM interaction
plays some role, the collision should generate
a double-shock pattern with the forward shock running into the
CSM and the reverse shock propagating back into the ejecta. 
Between the shocks is a contact discontinuity, where a cool, dense
shell (CDS) could form shortly after explosion. 
As the blast wave (the forward shock) runs through the dense
CSM around SN~2008am, the temperature behind the shock is high
enough to ionize both H and He. The emergent emission 
lines are thus expected to be due to radiative recombination.
The photons can then further interact with the CSM in the early phases,
resulting in the observed broadened emission lines.

The overall appearance of the 
intermediate-width Balmer emission lines
on a nearly featureless, blue continuum makes SN~2008am similar
to other Type IIn SNe, in particular the SLSN~2006gy (Smith et al. 2010).
Here we use this similarity to address the possible line-forming
mechanisms that may explain the observed spectral properties of
SN~2008am.

\subsection{{\it Possible Line-Forming Processes}}

Smith et al. (2010) delineated three phases for SN~2006gy.
In the first phase, between 0-90 days after explosion (extending
20 days after maximum), when SN~2006gy was within a factor of
2 of peak light, H$\alpha$ showed a nearly symmetric profile 
with no P-Cygni features. Smith et al. associate this phase
with conditions where there is a shock wave deep within a dense, opaque
circumstellar shell, but the photosphere is in the outer, unshocked 
CSM. The H$\alpha$ line is presumed to be excited by photoionization and
to be intrinsically narrow, but broadened by multiple scattering on
hot, free electrons (Fransson \& Chevalier 1989; Chugai, 2001). 
In the second phase, 90 - 150 days after explosion (20 - 80 days after maximum light),
the H$\alpha$ line in SN~2006gy broadened somewhat, and developed a distinct, narrow 
P Cygni component with a velocity of $\sim$200~km~s$^{-1}$ and strong absorption in the blue wing extending to 
$\sim$ 4000 km s$^{-1}$. The red wing is nearly constant during this phase
and indicates a line of intermediate width of $\sim$ 1800 km s$^{-1}$.
Smith et al. attribute this phase to a condition where the photosphere
has receded to beneath the forward shock so that radiation from the shocked matter
can escape freely. Smith et al. (2010) proposed that the width of the red wing is determined by the Doppler shift
of the expanding cold, dense shell (CDS) that is presumed to form between
the forward and reverse shocks in the collision of the SN ejecta with the
dense CSM (Chugai 2001). 
The third phase in SN~2006gy is the very late phase, 150-240 days after explosion, when
the H$\alpha$ line becomes narrower. This is the phase when the CSM interaction
is expected to decline in strength and the line-emitting region to become optically thin. 
SN~2008am and SN~2006gy had different rise times
in their light curves, but their spectral evolutions show similarities if 
account is taken for the different timescales. 

We first examine the possibility that the intermediate-width lines in the early phase
of SN~2008am obtain their 
wings from multiple electron scattering, a mechanism favored by Smith et al.
(2010) for SN~2006gy. Our first four spectra of SN~2008am were obtained when
the SN was within a factor of 2 of maximum light, in keeping with Phase 1
of SN~2006gy by Smith et al. Our spectra can be fit well with a single
Lorentzian profile (see below) consistent with multiple electron scattering,
and they show no sign of broad features nor P Cygni lines of any width, similar 
to Phase 1. For these reasons, we believe a plausible case can be made that, 
despite the different timescales, SN~2008am could be a close cousin to 
SN~2006gy, and that our early-phase spectra are formed in the phase when
the photosphere was in a dense CSM shell, but beyond the forward shock. If this is
the case, SN~2008am could very well have then proceeded to Phase 2 defined by
Smith et al. (2010), but we simply failed to obtain any spectra in this phase, more
than 2 magnitudes below maximum light.
At very late phases, the two objects displayed substantially narrower H$\alpha$ 
profiles and are again quite similar. Further support for this model may come from the fact that in the early phase both 
H$\alpha$ and H$\beta$ showed peaks redshifted by $\sim 100$ km s$^{-1}$ from
their rest-frame wavelength. The same effect was observed for SN~2006gy
by Smith et al. (2010) and it is also predicted by the electron scattering
model of Chugai (2001). Chugai (2001) assumed a velocity profile in which
the velocity decreased with radius as might be caused by radiative acceleration. Fransson \& Chevalier (1989)
assumed an homologous velocity profile in their multiple scattering models and found profiles with enhanced
red wings and a small blue shift of the peak. The details of the line profiles might
thus contain information on the velocity profiles in the scattering regions of
SN~2008am and related events, but we have not explored this in any depth.

The second possibility is that the intermediate-width lines arise from the
post-shock motion of the shocked CSM in a CDS 
when the photosphere is interior to the forward shock (Fransson 1984; Chugai et al. 2004;
Dessart et al. 2009). This would correspond to our early spectra being already in
the phase 2 of Smith et al. (2010).
The principal argument 
against this interpretation is that we do not see any of the narrow or
broad absorption manifested by SN~2006gy when it was in this phase, as interpreted
by Smith et al. (2010). It could be that the structure of the CSM 
around SN~2008am is such that the matter interior to and beyond the shock is
not sufficiently optically thick to create appreciable absorption. 
Another possibility is that these absorption features might have been observed had
we had better S/N ratio and/or better spectral resolution. We thus cannot rule out
this possibility, but find it somewhat less likely than the broadening by
multiple electron scattering. 

Another mechanism that has been proposed to account for intermediate-width lines
is the inward propagation of shocks into dense clumps of matter that
have been engulfed by the SN shock (Chugai \& Danziger, 1994), rather than a
single post-shock shell. In this picture, the SN shock sweeps past clumps in
the progenitor wind, but the resulting high pressure of the shocked low-density
CSM drives a shock into the dense clumps. This may broaden the lines via 
Doppler-motion, and for appropriate choices of cloud sizes, densities and other
parameters one can produce H$\alpha$ lines of suitable width and intensity.
The problem with this model in the current context is that it is designed to have
dense clumps separated by less dense, optically-thin material. The latter should
allow the SN ejecta to be observed directly. Since we see no sign of high-velocity 
SN features, we find the configuration with the enveloping CDS, as attributed
to SN~2006gy by Smith et al. (2010), to be preferable to the clumpy model, where
the dense clumps have a rather small filling factor.

\subsection{{\it Line profile fitting}}  

In the context of the CSM interaction picture, we considered 
three different models to account for the line profiles
of the early-phase spectra of SN~2008am: 
$i)$ thermal Doppler-broadening producing 
Gaussian line shapes, $ii)$ single Thompson scattering on free
electrons giving exponential profiles (Laor 2006) and  
$iii)$ multiple scattering on hot free electrons 
resulting in Lorentzian profiles (Chugai, 2001; Smith et al. 2010). 
The first phenomenon is common in stellar atmospheres. The 
second one is proposed for the broad-line region in AGNs/QSOs.
The third one might pertain to the dense CSM
environment around Type IIn SNe.

We fitted Gaussian, Lorentzian and exponential profiles
to the observed features. The results are listed in Table 8 and plotted
in Figure 12.
The Gaussian fits were inferior compared to the other two 
models in terms of conforming to profile shapes. Gaussian models resulted 
in FWHM = 1500 $\pm$ 300 km s$^{-1}$ for
H$\alpha$ and 1600 $\pm$ 500 km s$^{-1}$ for H$\beta$. 
If interpreted as a simple thermal Doppler-broadening, the corresponding
temperature would be $\sim 10^{8}$ K, which is obviously too high.
On the other hand, if the line width is attributed to bulk kinematic
motion, these velocities are too low to be directly related to 
the expected velocities of SN ejecta.

The exponential fits are motivated by the possibility that the 
line-forming region might be a photoionized, but less dense,
optically-thin medium, similar to the environment of AGNs.
In this case, the observed line profiles are due to Thompson
scattering on hot free electrons. If the medium is 
less dense, hence optically thin to electron scattering,
single scattering is an adequate description. Laor (2006) 
investigated such a region and derived
the emergent line shape to be $\sim exp(-\Delta v / \sigma)$, where
$\Delta v$ is the Doppler-shift from the line center 
and $\sigma$ is the velocity.  
The electron-scattering optical depth can be expressed as
$\tau_{e} = exp( - \sigma / (1.1 \sigma_e))^{2.222}$
where $\sigma_e \sim (k T_e / m_e)$ is the velocity
dispersion of free electrons. The fits to the observed
line profiles resulted in $\sigma$ = 800 $\pm$ 100 km s$^{-1}$,
which gives $\tau_e$ $\sim$ 0.03 if the electron
temperature $T_e$ is assumed to be 11,000 K, close to the
continuum effective temperature of SN~2008am. 
The resulting optical depth is much below unity,
verifying the general assumption of this model;
however, it should be noted that higher optical depths
produce narrower lines in this model. Since we
observe intermediate-width lines earlier and a narrower line later, the
prediction from the Laor model would be the strong
increase of electron optical depth, the opposite
of what is expected in an expanding, diluting SN environment.

The fitting of Lorentzians produced the best fit to all
line profiles. The resulting FWHMs are $\sim$ 1000 $\pm$ 300 km s$^{-1}$
for H$\alpha$ and $\sim$ 1700 $\pm$ 500 km s$^{-1}$
for H$\beta$. The narrower H$\alpha$ profile may be
the consequence of host contribution to the HET spectra,
which is certainly stronger in H$\alpha$. Indeed, the
FWHM of the host-subtracted early-phase Keck H$\alpha$ profile is much closer
to the average FWHM of the H$\beta$ line. Since the
line-forming medium is assumed to be optically thick
in this model, the line widths from
the Lorentzian fitting cannot be simply related to the
physical conditions characterizing the whole line-forming region.
A higher density of free electrons should
produce wider profiles and prevent the direct escape of the
original narrow-width line. As a consequence, as the line-forming 
medium expands, the line broadening should decrease. 

If multiple scattering is the correct interpretation, then the width of the lines we measure
in the early-phase spectra of SN~2008am cannot be attributed to bulk 
kinematic motion. The width of the lines gives an upper limit on the motion of the line-forming
region, but yields no constraint on the velocity of deeper, optically-thick regions. The profiles
may give information on the velocity distribution in the scattering region since the velocity profile can 
affect the resulting scattering-line profile if the bulk velocity exceeds the electron thermal velocity
(Fransson \& Chevalier 1989; Chugai 2001). 
For multiple scattering, the line wings are primarily a measure of the electron scattering
optical depth of the line-emitting region. Following Smith et al. (2010)
we estimate the Thompson scattering optical depth from the formula 
$U = (1 - e^{-\tau_T} )/ \tau_T$, where $U$ is the ratio of the narrow, unscattered
H$\alpha$ line flux to the total H$\alpha$ luminosity. Since we were unable to
fully resolve the narrow component in the +358d Keck spectrum, 
we use a conservative estimate of $U \leq 0.5$. This results in a lower limit of
$\tau_T \geq 2$ for SN~2008am (Smith et al. obtained $\tau_T \sim 15$ for SN~2006gy),
consistent with an optically-thick scattering medium.

\section{MODELS FOR SN~2008am}\label{mods}
 
Figure 13 shows the ROTSE unfiltered light-curve of SN~2008am compared to some other SLSNe and confirms that SN~2008am 
is one of the brightest SNe ever observed. Figure 13 also shows that the late photospheric evolution of SN~2008am 
(after $\sim$~120 d) in the R band is very similar to that of SN~2006tf and slower compared to all the other SLSNe except SN~2003ma (Rest et al. 2009). 
In the following discussion
we use the derived rise time and ROTSE light curve of SN~2008am (\S 2.2; Figure 3) to obtain estimates of the mass of involved ejecta+CSM
and to discuss possible power sources for this exceptional stellar explosion. 
To do so, we fit light curve
models that account for the diffusion of radiated energy deposited from a variety of power-sources. The fitting in each case
is performed by Monte-Carlo chi-square minimization as described in Chatzopoulos et al. (2009).
We also compare to our derived SED light curves, as appropriate.

\subsection{{\it Radioactive decay diffusion models}}\label{radif}

Although there is no evidence for a classic SN photosphere, the light curve of SN~2008am (Figure 3) is reminiscent of SNe
powered by radioactive decay.
 The first model that we consider for SN~2008am is thus one of radioactive decay diffusion that was developed by
Arnett (1980, 1982) and generalized by Valenti et al. (2008) (see also supplementary information in Soderberg et al. 2008). In this
model, the power source of the SN luminosity is the radioactive decay of nickel and cobalt, the energy of which
diffuses out from the expanding envelope. 
This model was developed in the context of
Type Ia SNe and is appropriate in the absence of a H recombination phase with a constant opacity in the photospheric
phase (the effects of H recombination were considered by Arnett \& Fu (1989) and Chatzopoulos et al. (2009)).
The light curve is given by the following
formula (Valenti et al. 2008; Chatzopoulos et al. 2009):
\begin{equation}
L(t)=M_{Ni}e^{-x^{2}}[(\epsilon_{Ni}-\epsilon_{Co}) \int_0^x2ze^{z^{2}-2zy}dz
  + \epsilon_{Co}\int_0^x2ze^{z^{2}-2yz+2zs}dz](1-e^{-At^{-2}}),
\end{equation}
where $x = t/t_{m}$, $t_{m}$ is the effective diffusion time which is generally close to the rise time to maximum,
$y = t_{m}/2t_{Ni}$ with $t_{Ni} =$~8.8 days,
$s = t_{m}(t_{Co}-t_{Ni})/2t_{Co}t_{Ni}$ with
$t_{Co} =$~111.3 days, $M_{Ni}$ is the initial nickel mass, and
$\epsilon_{Ni} = 3.9 \times 10^{10}$ erg~$~$s$^{-1}$~g$^{-1}$
and $\epsilon_{Co} = 6.8 \times 10^{9}$ erg~$~$s$^{-1}$~g$^{-1}$ are the energy generation rates
due to Ni and Co decay. The factor
$(1-e^{-At^{-2}})$ accounts for the gamma-ray leakage, where large $A$
means that practically all gamma rays are trapped. The gamma-ray optical depth of 
the ejecta is taken to be $\tau_{\gamma} = \kappa_{\gamma} \rho R = At^{-2}$, assuming spherical uniform density
ejecta with radius $R = vt$ and the Ni/Co confined in the center. 
This yields $A = (3\kappa_{\gamma} M_{ej})/(4 \pi v^{2})$ which is controlled by the
gamma-ray opacity, $\kappa_{\gamma}$. The $t^{-2}$ scaling follows from
homologous expansion, which is one of the basic assumptions of the simple analytic
models that we adopt here.
Thus the main parameters of this model are the nickel mass, $M_{Ni}$, and the effective diffusion time
$t_{m}$, which corresponds to an ejecta
mass as given by the following equation for a constant density envelope:
\begin{equation}
M_{ej}(z)=\frac{3}{10} \frac{\beta c}{\kappa} v \frac{t_{m,ob}^{2}}{(1+z)^{2}},
\end{equation}
where $\beta$ is an integration constant equal to about 13.8 (Arnett 1982;
Valenti et al. 2008), $\kappa$ is the mean optical opacity, $v$ the mean photospheric expansion velocity, $t_{m,ob}$ the observed
rise time and $z$ the redshift of the SN. As described in \S3.2 we have concluded that the observed emission lines are dominated
by multiple electron scattering so that their width yields no direct information on the bulk kinematic expansion velocity. 
In the following, we will scale our results to a characteristic velocity
of $v =$~1,000~km~s$^{-1}$ that is consistent with the limit set by line widths and might be compatible with velocities
expected for shocks traversing dense circumstellar shells.

We provide a characteristic fit of the radioactive decay diffusion model to the ROTSE light curve of SN~2008am
in order to estimate model parameters for this SN and to determine the applicability of this model.
 For the ROTSE light curve of SN~2008am the explosion date is held fixed, established in \S 2.2. Thus, we are
left with three fitting parameters: $M_{Ni}$, $t_{m}$ and $A$.  
The gamma-ray leakage in this model is very small over the range of our observations.
A decent fit is obtained for $M_{Ni} =$~19~$M_{\odot}$ and $t_{m} =$~41d.
The 41 day effective diffusion time corresponds to an ejecta
mass $M_{ej} =$~0.2~$M_{\odot}$, using Equation 2 with the fiducial values $\kappa =$~0.4~$cm^{2}~g^{-1}$, appropriate for Thompson scattering in
a pure H plasma, and $v =$~1,000~km~s$^{-1}$. If we use $v =$~2,000~km~s$^{-1}$ instead, which would be above the upper limit for the velocity implied from
the observed FWHM of the emission lines in the early spectra of SN~2008am, and adopt a lower value for the
optical opacity, $\kappa =$~0.05~cm$^{2}$~g$^{-1}$, characteristic of metal-rich ejecta, the ejecta mass can be scaled up to 3.2~$M_{\odot}$.
Even for this low opacity, the velocity would have to be $v \sim$~12,000~km~s$^{-1}$ for the diffusion mass to exceed the required nickel
mass, and we have no rationale to adopt such a high velocity.

The left panel of Figure 14 shows the radioactive decay diffusion model (red solid curve) fitted to the ROTSE light curve of SN~2008am 
(blue filled squares). The radioactive decay rate of cobalt for $M_{Ni} =$~19~$M_{\odot}$ is shown as the red dashed curve in the same plot for comparison.
The light blue data points correspond to the pseudo-bolometric light curve of SN~2008am derived by the SED fits. 
As we have shown above, the value of $M_{ej}$ would be greater for lower $\kappa$ and higher $v$, 
but it never becomes equal to or higher than the nickel mass for reasonable choices of these parameters
given the short rise-time of SN~2008am. This makes the radioactive decay diffusion model
unphysical for this event.
Although a large amount of radioactive nickel has been suggested for some SLSNe as a product of pair-instability (for example, 22~$M_{\odot}$
for SN~2006gy; Smith et al. 2007 and 4.5~$M_{\odot}$ for SN~2006tf; Smith et al. 2008) the inconsistency between the total ejected mass 
and the nickel mass in SN~2008am is quite
remarkable. Integrating under the solid red curve of the left panel of Figure 14 yields a total 
radiated energy output $E_{rad} \simeq 2.1 \times 10^{51}$~erg.
On the other hand, the kinetic energy of the explosion in this model, $E_{KE} = (1/2) M_{ej} v_{mean}^{2}$ 
is $0.6 \times 10^{48}$~erg where $v_{mean} = \sqrt{3/5} v$
assuming $v =$~1,000~km~s$^{-1}$ and for $\kappa =$~0.4~$cm^{2}~g^{-1}$. Even within the uncertainties of these estimates, the kinetic
and total radiated energy are also found to be inconsistent. 
Radioactive decay may contribute to the output 
energy, but the most significant contribution must come from other mechanisms. We discuss other models in the following
sections. 

\subsection{{\it Shell-shock diffusion model}}\label{comp}

 Here we consider a shell-shock diffusion model, similar to that suggested by Smith \& McCray (2007) for SN~2006gy. In this type of model
the energy that powers the SN light curve
is produced by the diffusion of shock-generated energy through an optically-thick CSM shell of large initial radius. 
Smith \& McCray (2007) considered the diffusion of shock-generated energy in a homologously expanding CSM shell in the case where 
the shock energy input is instantaneous and at the time of maximum light. This type of model does not account for the observed rise of
the light curve. 
Smith \& McCray (2007) adopted a L$\propto r^{2}$ rise for their model of SN~2006gy based on the early portion of the diffusion models of Arnett 
(1982). They did not self-consistently consider the input necessary to drive such a rise. They found the diffusion time on the decline to be about
the observed rise time for SN~2006gy, but we consider that a coincidence, since such a model fails to account for SN~2008am, as we show below.
In order to account for the rising part of the light
curve of SN~2008am, we consider a forward shock that propagates through the CSM envelope and deposits energy for a time $t_{sh} = t_{max}$ and then shuts off.
The effect of the reverse shock will be considered below.
This model is somewhat similar to the ``top hat" magnetar-input model that is considered in Kasen \& Bildsten (2010) with the exception that
we solve for the general case of large initial radius. The luminosity deposition function in our model has the form $L_{sh}(t) = E_{sh}/t_{sh}$ for $t<t_{sh}$
and $L_{sh} = 0$ otherwise, where $E_{sh}$ is the total kinetic energy deposited by the shock in
the CSM shell.
This model formally assumes that the luminosity $E_{sh}/t_{sh}$ is deposited in the center of homologously expanding matter. While not
totally self-consistent, this model captures the essence of the shell-shock model on both the rise and decline.
Using this energy deposition function and the first law of thermodynamics coupled with the diffusion approximation
(as was originally done by Arnett 1980, but for a radioactive decay input), it can be shown that the light curve 
is given by the following expression:
\begin{eqnarray}
L(t) = 
\cases{\frac{E_{sh}}{t_{sh}} [1-e^{-(t^{2}/2t_{d}^2+R_{0}t/v t_{d}^{2})}], & $t < t_{sh}$, \cr
\frac{E_{sh}}{t_{sh}} e^{-(t^{2}/2t_{d}^2+R_{0}t/v t_{d}^{2})} [e^{(t_{sh}^{2}/2t_{d}^2+R_{0}t_{sh}/v t_{d}^{2})}-1], & $t > t_{sh}$,}
\end{eqnarray}
where $t_{d}$ is the diffusion time-scale, $v$ is the characteristic bulk velocity of 
the CSM shell, and $R_{0}$ is the initial radius of the optically-thick CSM shell around
SN~2008am. We note that in the case where $R_{0}$ is small,
the result for small initial radius is recovered (Kasen \& Bildsten 2010).

To evaluate this model, the second part of Equation 3 ($t > t_{sh}$) is fitted to the ROTSE light curve decline of SN~2008am 
in order to determine $E_{sh}$, $R_{0}$ and $t_{d}$.
Then using the best-fitting parameters, we plot the expected rise to maximum for the event within this class of model. We first consider the case
for which the diffusion time on the decline is forced to be equal to the input time, which is also the rise time to maximum ($t_{max} =$~34d).
The best fit to the data in this case, is obtained for $E_{sh} = 1.6 \times 10^{51}$~ergs and
$R_{0} = 3.0 \times 10^{13}$~cm if $R_{0}$ is constrained to be larger than the smallest possible progenitor radius ($10^{11}$~cm for Wolf-Rayet stars).
Forcing $t_{d}$ on the decline to be equal to the rise time of 34d means the light curve falls below
the indicated lower limits. 
The result of this failed ``fit" is shown as the red dashed curve in the middle panel of Figure 14.
We next consider the general case where $t_{d}>t_{sh}$, and we also fit $t_{d}$ as a free parameter. 
The best fit is obtained for $E_{sh} = 5.5 \times 10^{51}$~ergs,
$R_{0} = 1.0 \times 10^{14}$~cm and $t_{d} =$~87d. The result of this fit is 
shown as the red solid curve in the middle panel of Figure 14.
The 87d diffusion time corresponds to a CSM shell having a mass
$M_{shell} =$~1.0~$M_{\odot}$ (using Equation 2, again for $v =$~1,000~km~s$^{-1}$ and $\kappa =$~0.4~$cm^{2}~g^{-1}$). 
For the extreme values of $v=$~2,000~km~s$^{-1}$
and $\kappa =$~0.05~$cm^{2}~g^{-1}$ the shell mass is 16~$M_{\odot}$. The radius of the shell at maximum light adopting a
constant expansion velocity equal to 1,000~km~s$^{-1}$ is $R_{max} = 3.7 \times 10^{14}$~cm.
These results are consistent with a relatively 
large optically-thick CSM shell around SN~2008am. The total radiated energy within the context of this model is equal to $10^{51}$~erg.
Note that the rise in this simple ``top hat" model is concave upward rather than convex.

We conclude that the derived parameters for the CSM shell of SN~2008am based on a shell-shock diffusion model show that it may be
less massive and smaller than the ones determined
for SN~2006gy ($M_{shell}=$~10~$M_{\odot}$, $R_{shell} =$~$2.4 \times 10^{15}$~cm; Smith \& McCray 2007) and for SN~2006tf 
($M_{shell} =$~18~$M_{\odot}$, $R_{shell} =$~$2.7 \times 10^{15}$~cm, Smith et al. 2008 using the same type of model). 
It should be noted that the estimates for the properties of the shell
of SN~2006tf are very uncertain due to the lack of data during the rising part of the light curve and thus the lack of an accurate
explosion date. The larger mass derived for SN~2006gy is mainly determined by the larger rise time of 70d versus 34d for SN~2008am and by the
larger velocity of 4,000~km~s$^{-1}$ versus the fiducial 1,000~km~s$^{-1}$ that we have adopted for SN~2008am.

An estimate of the optical depth of the CSM shell under the assumption that the derived shell radius is significantly larger
than the radius of the progenitor and for constant density profile is given by:
\begin{equation}
\tau_{shell}=\frac{3 \kappa M_{shell}}{4 \pi R_{shell}^{2}}.
\end{equation}
Assuming that electron scattering in a hydrogen plasma is the main source of opacity ($\kappa =$~0.4~$cm^{2}~g^{-1}$),
$v=$~1,000~km~s$^{-1}$ (and taking for the CSM shell the derived parameters $M_{shell} =$~1.0~$M_{\odot}$ and 
$R_{shell} = R_{max} = 3.7 \times 10^{14}$~cm)
an estimate for the 
optical depth of the shell around SN~2008am at maximum light is $\tau_{max} \sim$~1,390 while for SN~2006tf it is $\tau_{max} \sim$~480 (Smith et al. 2008)
and for SN~2006gy it is 
$\tau_{max} \sim$~330 (Smith et al. 2007). Thus, although less massive, the CSM shell of SN~2008am is very optically thick due to its small radius
compared to the shells of those other SLSNe, which results in higher density.
The values of the
optical depth will vary for different choices of optical opacity. For the derived optical depth, we can estimate when the shell will
become optically thin.
Since $\tau \propto R^{-2}$, $R_{2} = R_{max} \sqrt{\tau_{max}/\tau_{2}}$ for a constant expansion velocity, where
$R_{2} = R_{max}+v\Delta t$. Combining those two equations yields $\Delta t = R_{max} (\sqrt{\tau_{max}/\tau_{2}}-1)/v$. Thus we can estimate
how long will it take for the CSM to become optically-thin ($\tau_{2}=1$) for $\tau_{max} =$~3,600 and $R_{max} = 3.7 \times 10^{14}$~cm. 
This yields $\Delta t \simeq 4.3$~yr for $v =$~1,000~km~s$^{-1}$. 

An estimate for the mass-loss rate is given by $\dot{M} = M_{shell} v_{w}/(t_{max} v)$. For SN~2008am ($M_{shell} =$~1.0~$M_{\odot}$), 
we find $\dot{M} = 0.9 \times 10^{-2} v_{w}$~$M_{\odot}$~yr$^{-1}$ for $v =$~1,000~km~s$^{-1}$ and with $v_{w}$ in units of km~s$^{-1}$. 
For the range of wind velocities 10~km~s$^{-1}$ $< v_{w} <$~1,000~km~s$^{-1}$ the inferred mass loss rates range from 
0.1 up to 10~$M_{\odot}$~yr$^{-1}$. These extraordinary mass loss rate estimates imply a very massive LBV-type progenitor for SN~2008am, 
as was suggested for SN~2006gy (Smith et al. 2008). 
We note again that the choice of 1,000~km~s$^{-1}$ as the expansion velocity of the circumstellar shell, which was constrained by 
fitting Lorentzian profiles to
the Balmer emission lines in the early spectra of SN~2008am (see \S 3.2),
is just for illustration purposes.

We also investigated the possibility that SN~2008am is
powered by the reverse shock that we expect to have formed due to an ejecta-CSM interaction. 
Chevalier \& Fransson (2003)
estimate the density of the swept up matter behind the reverse shock to be
\begin{equation}
\rho_{rev}=\frac{(-n-4)(-n-3)}{2} \rho_{CSM},
\end{equation}
where $n$ is a constant that describes the density profile of the ejecta ($\rho_{ej} \propto R^{n}$). The density behind the reverse
shock is higher than that behind the forward shock for $n \leq -7$. 
The luminosity from the reverse shock scales as $L_{rev} \propto t^{-3/(-n-2)}$. The reverse shock may be adiabatic or radiative 
depending on the optical depth of the shell.
We fit a power law  of the form $L = A t^{p}$
to the decline of the observed ROTSE light curve
in order to get an estimate of $n$ and determine whether this model can account for SN~2008am. 
This procedure effectively assumes that the reverse shock dominates the forward shock and that the diffusion time in the
reverse-shocked gas is small.
The best fit power law model gives $p = -1.1 \pm 0.3$ which corresponds to 
$n = -4.7 \pm 0.1$. Since $n>-7$ we conclude that the density behind the reverse shock may be smaller than that behind the forward shock 
and hence that the luminosity from the reverse shock may be lower than that of the forward shock.

\subsection{{\it Optically-thin CSM-ejecta interaction}}\label{csm}

A related model for the luminosity source for SN~2006gy (Ofek et al. 2007), SN~2002ic (Hamuy et al. 2003) 
and SN~2005gj (Aldering et al. 2006) 
is that of the interaction between the ejecta and the CSM in the case where the latter is dense, but optically-thin. In such a model, a radiative
shock forms due to the collision between the ejecta and the CSM (Chevalier \& Fransson 1994). We consider the luminosity
produced by the forward shock.
The luminosity in this case given by (Ofek et al. 2007):
\begin{equation}
L = 2 \pi \rho_{CSM} R^{2} v_{sh}^{3},
\end{equation}
where $\rho_{CSM}$ is the local CSM density, $R = v_{sh} t$ is the radius of the shock at time $t$ and $v_{sh}$ the velocity of the shock. Assuming
that the shock enters the optically-thin CSM envelope at a radius $R_{0}$ and that the density profile follows a power law of the form $\rho =
\rho_{0} (R/R_{0})^{m}$, where $R = R_{0}+v_{sh} t$ and $m$ is the slope of the density profile (for a constant velocity wind $m$=-2), we
can re-write Equation 6 as: 
\begin{equation}
L = 2 \pi \rho_{0} R^{2} v_{sh}^{3} [(R_{0}+v_{sh}t)/R_{0}]^{m}.
\end{equation}
In our analysis, we assume that the shock velocity
is constant for simplicity.  
More accurate solutions that take the change of the shock velocity into account can be found in Chevalier (1982) 
and Chevalier \& Fransson (2003).
If the CSM around the progenitor star is everywhere optically-thin, $R_{0}$ represents the radius of the progenitor star. 

An estimate for an upper limit of the mass-loss rate of the progenitor of SN~2008am in the context of an optically-thin
CSM comes from the X-ray flux upper limit
measurement (see \S 2.2). According to Immler \& Kuntz (2005), the X-ray luminosity from thermal Bremsstrahlung is given by:
\begin{equation}
L_{X}=\frac{4}{(\pi m)^{2}} \Lambda(T) (\frac{\dot{M}}{v_{w}})^{2} (v_{sh} t)^{-1},
\end{equation}
where $L_{X}$ is the X-ray luminosity, $m_{p}$ the mean mass per particle (we adopt $m_{p} = 2.1 \times 10^{-24}$~$g$ for a H+He plasma) and $\Lambda(T)$ the cooling function
of a plasma that has temperature $T$. As in Immler \& Kuntz (2005) we adopt $\Lambda(T) = 3 \times 10^{-23}$~erg~cm$^{-3}$~s$^{-1}$ for $T = 10^{7}$~K which is 
characteristic of the post-shock temperature. For the available XRT upper limit of $L_{X} = 10^{43}$~erg~s$^{-1}$ corresponding 
to $\sim$~50 rest frame days after explosion and 16 days after maximum, we obtain an upper limit of
$\dot{M} = 0.07$~$(v_{w}/1,000$~km~s$^{-1})$~$M_{\odot}$~yr$^{-1}$. It should be noted that this model assumes a constant
wind-density parameter $w = \dot{M}/v_{w}$.

Most of the radiation that is produced in models of the radiative forward shock is emitted at ultraviolet and X-ray wavelengths.
We assume in the context of this model that
the ROTSE light curve is a good proxy to the true bolometric light curve of SN~2008am as would be the case if a substantial
fraction of UV/X-ray luminosity is absorbed and re-radiated by the shell that forms behind the forward shock.
Then, the best fit of Equation 7 to the ROTSE light curve of SN~2008am would provide us with estimates of $\rho_{0}$, $R_{0}$ and $m$. 

The optical depth of the CSM is
\begin{equation}
\tau_{thin} = \kappa \int_{R_{0}}^{\infty} \rho dR = -\frac{\kappa \rho_{0} R_{0}}{m+1},
\end{equation}
where $m<-1$. In our fitting process we demand that this optical depth beyond $R_{0}$ is less than one for $\kappa =$~0.4~$cm^{2}~g^{-1}$.
Thus we choose the best-fit model for which this condition is met in order to be self-consistent with the initial assumption of an optically-thin CSM.
We were unable to determine a satisfactory fit of this simple optically-thin model to the ROTSE light curve of SN~2008am that was self-consistent, 
in terms of the optical depth.
For all the best fit models the optical depth was well above unity, and for models with $\tau_{thin}<1$ the derived values for the density and the radius
where unphysical.

A purely optically-thin model not only fails to account for the light curve of SN~2008am but also fails to account for
the observed spectroscopic characteristics. Chevalier (1982) and Chevalier \& Fransson (2003) predict that the bulk of the luminosity from optically
thin forward shock emission is in the UV and X-ray region of the spectrum. It seems unlikely that the optically-thin case would reproduce
the spectra and spectral evolution of SN~2008am.
In addition, this model being optically thin, it is incompatible with the failure to see SN photospheric features.
We conclude that an optically-thin ejecta-CSM interaction model alone does not provide a satisfactory explanation for SN~2008am.

\subsection{{\it Hybrid CSM Interaction Model}}

 Next, we discuss a hybrid model in which the CSM comprises both an optically-thick and an optically-thin region. This is somewhat similar
to the model proposed by Chugai \& Danziger (1994) in which the CSM contains two components: an optically-thin rarefied wind and optically-thick 
clumps or an optically-thick disk around the progenitor. In this model, the lightcurve
is described by the sum of Equation 3 and Equation 7. We assume the whole
structure follows a single power-law density profile $\rho \propto R^{m}$ so that the optical depth at a given point is:
\begin{equation}
\tau = \frac{-\kappa \rho_{0} R}{m+1} (\frac{R}{R_{0}})^{m}.
\end{equation}
We take the point $\tau =$~1 at radius $R = R_{1}$ to represent the boundary between the optically-thick and optically thin regions. In
this case, $R_{0}$ corresponds to the inner initial radius of the optically-thick part of the CSM shell, which is also the radius
of the progenitor star, where the density is $\rho_{0}$.
The radius of unity optical depth
is $R_{1} = [-R_{0}^{m} (m+1)/ (\kappa \rho_{0})]^{1/(m+1)}$,
and we demand that this radius
is larger than $R_{0}$ so that the CSM has both an optically-thin and an optically-thick part. This requires a slightly different
set of fitting parameters: $R_{0}$, $t_{sh}$, $t_{d}$, $m$, $\rho_{0}$ and $E_{sh}$, and we again fix $t_{sh} = t_{max}=$~34d.
The best fit (plotted in the middle panel of Figure 14) is the same as that obtained in the purely optically thick shell case 
indicating that any contribution from optically thin emission is negligible.
Therefore, for a hybrid model, the optically-thick component (Equation 3) dominates all phases of the observed ROTSE light curve of SN~2008am. 

\subsection{{\it A magnetar powered SN~2008am?}}\label{csm}

 Kasen \& Bildsten (2010) and Woosley (2010) proposed the idea that the light curves of some SLSNe may be powered by the spin-down of 
young magnetars. In such a model, the energy input by the magnetar
is given by the dipole spin-down formula:
\begin{equation}
L_{p}(t) = \frac{E_{p}}{t_{p}} \frac{l-1}{(1+t/t_{p})^{l}},
\end{equation}
where $E_{p}$ is the initial magnetar rotational energy, $t_{p}$ is the characteristic time scale for spin-down that depends on the strength
of the magnetic field and $l=$~2 for a magnetic dipole. For a fiducial moment of inertia, the initial
period of the magnetar in units of 10~ms is given by $P_{10} = (2 \times 10^{50}$erg~s$^{-1}$ /$E_{p})^{0.5}$. The magnetic field of the magnetar can be estimated
from $P_{10}$ and $t_{p}$ as $B_{14} = (1.3 P_{10}^{2}/t_{p,yr})^{0.5}$, where $B_{14}$ is the magnetic field in units of
$10^{14}$~G and $t_{p,yr}$ is the characteristic time scale for spin-down in units of years.

Adopting Equation 11 as the energy deposition function (instead of the corresponding one for the radioactive decays of Nickel and Cobalt that leads
to the Arnett (1980) solution) and using the first law of thermodynamics coupled with the diffusion approximation, it can be shown that
the light curve of a supernova powered by a magnetar is given by the following expression:
\begin{equation}
L(t)=\frac{E_{p}}{t_{p}} e^{-x^{2}/2} \int_{0}^{x}e^{z^{2}/2} \frac{z}{(1+yz)^{2}}dz,
\end{equation}
where $x = t/t_{d}$ and $y = t_{d}/t_{p}$ with $t_{d}$ being the characteristic diffusion time.
In this treatment we assume that the initial radius of the progenitor star, $R_{0}$ is small.
We fit Equation 12 to the ROTSE light curve of SN~2008am and we obtain $E_{p} = 3.2 \times 10^{51}$~erg,
$t_{d} =$~64~d and $t_{p} =$~64~d. 
Note that in this model, neither $t_{p}$ nor $t_{d}$ corresponds to the time of maximum light.
The fit of this model is shown as the red solid curve in the right panel of Figure 14. Using the fitted
values of $E_{p}$ and $t_{p}$, the implied initial period of the magnetar would be 2.5~ms and the magnetic field 
$\simeq 0.7 \times 10^{14}$G. The total radiated energy implied by the magnetar model is $\sim 10^{51}$~erg.
The values of $P$ and $B$ lie within the range predicted in Duncan and Thompson (1992) for magnetars assuming the field to arise
in an $\alpha$-$\Omega$ dynamo. The magnetar model does provide a reasonable fit to the data including the
rise time (Figure 14; right panel). Although the magnetar model
provides a decent fit to the light curve, it gives no natural explanation for the emission lines.

\section{CONCLUSIONS}\label{disc}

We presented an analysis of the available photometric and spectroscopic data of the SLSN 2008am. 
The spectroscopic signatures of intermediate width H and He emission lines ($\sim$~25 \AA) place 
this SN in the category of Type IIn. SN~2008am was an extremely luminous event, 
with a peak absolute R-magnitude of $M_{R} \simeq -$22.3~mag corresponding to a luminosity of 
$\sim 2\times10^{44}$ erg s$^{-1}$, putting it in the ``hall of fame" of the most luminous 
SNe ever observed. The host of SN~2008am is a faint extended galaxy with 
magnitude $\sim$~20 in the SDSS catalog. At the redshift of $z =$~0.2338, the 
absolute r$^{\prime}$-magnitude of the host is $M_{r^{\prime}} \sim -$20~mag, which is in the 
range typical for elliptical and spiral galaxies. The very late (+554d) Keck spectrum of 
SN~2008am is consistent with an SB1 template spectrum. From the line flux ratios
the host has sub-solar metallicity, $Z \sim $~0.4~$Z_{\odot}$.  
We conclude that the host of SN~2008am is a metal-poor, but normal galaxy, 
not a subluminous dwarf as is the case for many SLSNe (Miller et al. 2009;
Drake et al. 2010; Miller et al. 2010).

SN~2008am was followed up photometrically from the IR to the UV. The ROTSE light curve
provides a reasonably accurate estimate of the rise time to maximum light
of 34 d in the rest frame, a significantly short time compared to other SLSNe. The 
photometric coverage allowed us to create broad-band SEDs of SN~2008am for 6 epochs. 
We fit single temperature black-body curves to the SEDs to study the evolution of the 
black-body temperature and radius as well as to estimate a pseudo-bolometric light curve.  
The derived black-body temperatures ($\sim$~10,000-12,000~K) 
are consistent with the continua of contemporaneous spectra.
These temperatures are very high compared with typical core-collapse supernova photospheres (5,000-6,000 K) as well as with the 
temperatures obtained for SN~2006gy and SN~2006tf (6,000-8,000 K; Smith et al. 2008, 2010). 
The single-temperature black-body fits for SN~2008am were imperfect, implying that 
the underlying emission mechanism is more complex in nature. 


Spectra obtained about 10 - 30 rest-frame days after maximum light showed intermediate-width 
emission lines of H$\alpha$, H$\beta$, H$\gamma$, and a feature at the HeI/Na D blend. There 
is no sign in our data of broad P Cygni lines that might signify the photosphere of the 
underlying supernova nor of narrow P Cygni lines as displayed by SN~2006gy and SN~2008tf 
on decline that indicate absorption in the unshocked, but expanding circumstellar matter
(Smith et al. 2008, 2010).  A spectrum obtained 352 days after 
our estimated maximum showed only H$\alpha$ that was significantly narrower than in the earlier 
spectra. Our failure to detect P Cygni features is most likely attributable to our lack of 
data at phases fainter than 2 magnitudes from maximum when SN~2006gy began to show such features, 
although we cannot rule out issues of 
S/N ratio and wavelength resolution.  The overall observed spectral evolution of SN~2008am
is similar to that of SN~2006gy at similar phases, and we conclude that they are closely related.

We considered a variety of models for the emission line profiles: Gaussian as might 
typify thermal Doppler broadening, an exponential profile that might characterize
single electron scattering in an optically-thin medium, and Lorentzian that might represent
models of multiple electron scattering (Chugai 2001; Smith et al. 2010). We find that
the latter provides the best fit to the overall line shape. An important implication
is that the line broadening is probably dominated by electron scattering in our 
spectra and that the line width contains little or no information about the bulk kinetic expansion 
velocity of the matter in the circumstellar medium or the underlying supernova. An upper 
limit to the velocity in the line-forming region is about 1,000 km s$^{-1}$. Chugai made
specific assumptions in his models, for instance that the velocity profile
decreased outward as might result from radiative acceleration, that might
not apply in general. The line profiles might contain information about the velocity structure even if
scattering dominated (Fransson \& Chevalier 1989). The radiative transfer that results in these line profiles
is worthy of re-examination.
  

We explored a number of light curve models based on a 
generalization of the models of Arnett (1980; 1982) that use a specified power input
and the first law of thermodynamics coupled with the diffusion equation. The models
are constrained by the rise time, the quasi-bolometric light curve and the general nature of 
the emission line spectra that show intermediate-width lines with little or no sign of 
P Cygni absorption on any scale. We examined models based on radioactive decay, an 
underlying supernova striking an optically thick circumstellar shell, a supernova shock 
in an optically-thin CSM, and a magnetar. The radioactive decay model is ruled out
because the required nickel mass would exceed the deduced ejecta mass. The magnetar model
provides a decent fit to the light curve, but no natural explanation for the emission-line structure. 
A shell-shock model similar to that of Smith \& McCray (2007) for SN~2006gy
in which the diffusion times on the rise and the decline are equal fails drastically for 
SN~2008am, primarily because of the rapid rise, 34 d in the rest frame. We conclude that 
the success of a model with a single diffusion time for SN~2006gy was a coincidence and 
not a general property of this class of events. We could generate a reasonable fit 
in the context of this model by including an input power source representing the 
collision of the supernova with an optically-thick shell that was a ``top hat" function of constant 
luminosity during the rise that shut off at 34 days, the point of maximum light in this
model. 
In this model, the timescale of the power source dictated the rise time and a 
separately determined diffusion time of 120d governed the decline. This model rose with 
$L \propto t^2$ at very early times (as assumed for the full rise of SN~2006gy by 
Smith \& McCray) and was overall concave upward in contrast to the observed light curve
that appears to be concave downward. The shape of the rise depends on the input power profile. We 
will investigate more general models in a future paper.  The optically-thin model is not 
an Arnett-like diffusion model, but assumes that the shock energy is rapidly radiated. 
This model did not provide a self-consistent fit to SN~2008am and gives no natural explanation for the 
rise nor for the failure to see any high-velocity features corresponding to the 
photosphere of the underlying supernova. 

The best fit to the ROTSE light curve data was obtained with an ejecta-CSM interaction model 
in which the supernova is not seen directly and the luminosity is produced by shocks and diffusion in a 
circumstellar medium that is optically thick. For fiducial parameters of $v = 1,000$~km~s$^{-1}$ and 
$\kappa = 0.4$~cm$^{2}$~g$^{-1}$, this model gives a rather small initial radius for the
shell, $R_{0} \sim 1.0 \times 10^{14}$ cm, with a mass of about 1~M$_{\odot}$ and a total energy input
from the underlying SN/CSM shock of $E_{s} = 5.5 \times 10^{51}$ erg.  
The model suggests that the optically-thick component may dominate 
the luminosity and that the forward shock provides a greater contribution to the luminosity 
than the reverse shock. 
This model suggests a rather large mass loss rate for the progenitor,
as perhaps would be consistent with an LBV-type mass loss process.

As noted above, SN~2008am seems to be a close cousin of SN~2006gy, showing Lorentzian 
emission lines shortly after maximum and a narrower H$\alpha$ line about a year later. 
The Lorentzian lines in both events show a slight redshift of about 100 km s$^{-1}$. The 
most notable difference is our failure to see the distinct broader (4,000 km s$^{-1}$)
and narrow (200 km s$^{-1}$) P Cygni features that appeared in SN~2006gy 20 to 80
days after maximum light. Smith et al. (2010) attribute the first phase of pure emission
to conditions where the shock is still beneath the photosphere of the dense CSM and the second
phase where P Cygni features form to conditions where the shock has proceeded beyond 
the photosphere. The high-velocity absorption is presumably related to the SN ejecta, 
and the low-velocity absorption to the motion of the CSM that has not yet been hit by 
the shock, but is subject to radiative excitation and recombination. The most likely 
explanation of our failure to detect these P Cygni features is absence of data at the 
appropriate phase. The narrower H$\alpha$ line in both events nearly a year after 
explosion is consistent with the emitting matter becoming more dilute with less 
broadening by multiple electron scattering.  

Other SLSNe seem to fall broadly in the category of SN~2006gy and SN~2008am. 
SN~2006tf shows nearly symmetric emission lines, especially of H$\alpha$, up
to 40 days after discovery (there is no data on the rise so the explosion date
and date of maximum are uncertain), with the development of narrow P Cygni features
by 66 days after discovery (Smith et al. 2008). At these later epochs, there are indications 
in both the emission and absorption for rapidly moving material, $\sim 7,500$ km s$^{-1}$,
presumably from the underlying SN ejecta. SN~2006tf is somewhat different
from SN~2006gy and SN~2008am in the late phases, a year after explosion, where
the H$\alpha$ line seems to be formed by collisional rather than radiative
excitation (Smith et al. 2008). The H$\alpha$ line in SN~2006tf at this stage shows a prominent blue 
``plateau" extending to about 1,000~km~s$^{-1}$. SN~2008iy was an unusual SN IIn 
with the unprecedented slow rise time of 400 days (Miller et al. 2010). The 
post-maximum spectra of SN~2008iy are somewhat similar to those of SN~2008am, with strong 
intermediate-width H and He emission lines. A single temperature black body failed to 
provide a good fit to the SEDs of SN~2008iy, as we found for the SED of SN~2008am (\S 2.3).
Miller et al. identified three distinct components in the late-time H$\alpha$ profile 
of SN~2008iy: broad ($\sim$~4,500~km~s$^{-1}$), intermediate ($\sim$~1,650~km~s$^{-1}$) 
and narrow ($\sim$~75~km~s$^{-1}$). Miller et al. proposed a model of interaction of 
the SN ejecta with a clumpy circumstellar medium similar to that for 
SN~1988Z presented by Chugai \& Danziger (1994). Miller et al. argued that the rise in 
the light curve resulted from an increase in the number of clumps with radius. 
Given the success of the shell-shock models of SN~2006gy and SN~2008am, 
it would be of interest to apply such a model to SN~2008iy.

As for SN~2006gy and SN~2006tf, the suggestion of a relatively massive shell around 
SN~2008am and the estimated mass loss rates ($\sim$~0.1-10~$M_{\odot}$~yr$^{-1}$)
imply that the progenitor star must have undergone substantial mass-loss in the 
years prior to the explosion.  
Episodic mass loss can occur around very massive LBV stars, similar to $\eta$ Carinae. 
The reason for LBV mass-loss is not currently fully understood (Smith \& Owocki 2006).
Massive shell ejection can also be the product of pulsational pair-instability 
(Rakavy \& Shaviv 1967; Barkat, Rakavy \& Sack 1967). Models suggest that repetitive
shell ejection takes place for progenitor main sequence masses in the 
range 95-130~$M_{\odot}$ (Woosley, Blinnikov \& Heger 2007).  
Supernova-like luminosity can be produced either during the ejection of each of these shells 
individually (Kasen et al. 2008), or during the collisions between shells ejected at 
different times (Woosley, Blinnikov \& Heger 2007). 

Although there is some sign of high-velocity material in some of the SLSNe that otherwise 
resemble SN~2008am, the nature of the presumed underlying supernova in these Type IIn
SLSNe remains obscure. As noted in the Introduction, other SLSNe show little or no evidence 
for hydrogen or interaction with a CSM. In the case of SN~2007bi, this is an important part 
of the argument that it is a pair-instability supernova (Gal-Yam et al. 2009).  
An important goal in the study of SLSNe remains the determination of the density
distribution in the CSM that will give important clues to the mass-loss history.

We are grateful to the anonymous referee for valuable guidance on style and science and 
to Andy Howell and Milos Milosavljevic for useful discussions. This research is supported 
in part by NSF Grant AST-0707669 and by the Texas Advanced Research Program grant ASTRO-ARP-0094. 
E. Chatzopoulos would like to thank the Propondis foundation of Piraeus, Greece for its support 
of his studies. J. Vinko received support from Hungarian OTKA Grant K76816.


{}                           


\begin{figure}
\begin{center}
\includegraphics[angle=0,width=15cm]{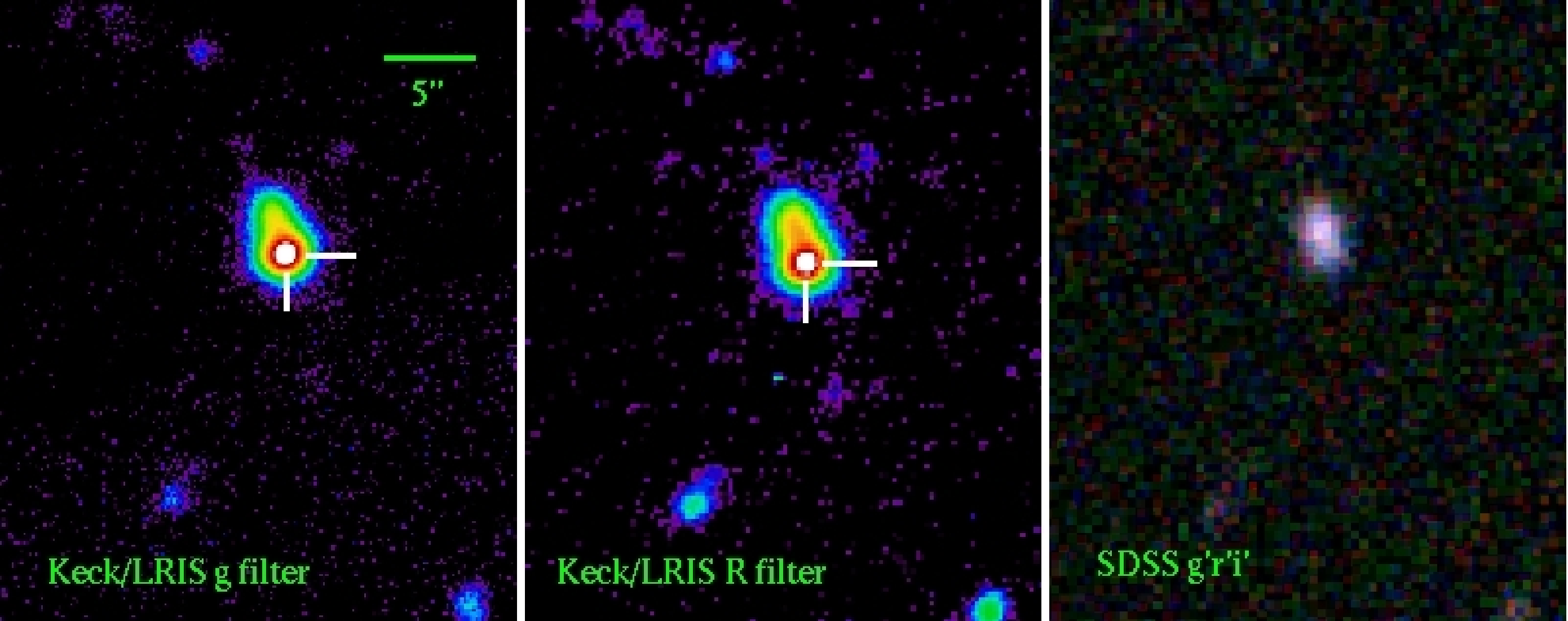}
\caption{False-color images showing the position of SN 2008am relative to its
host galaxy in the Keck/LRIS $g$ (left panel) and $R$ (middle panel) filters. Both Keck images were
taken on 2008 Feb 12.
The right panel contains the true-color SDSS image of the host (blue=$~g^{\prime}$, green=$~r^{\prime}$, red=$~i^{\prime}$ filters). 
The scale is indicated in the left panel. North is up and East is to the left on all images.}
\end{center}
\end{figure}

\begin{figure}
\begin{center}
\includegraphics[angle=270,width=15cm]{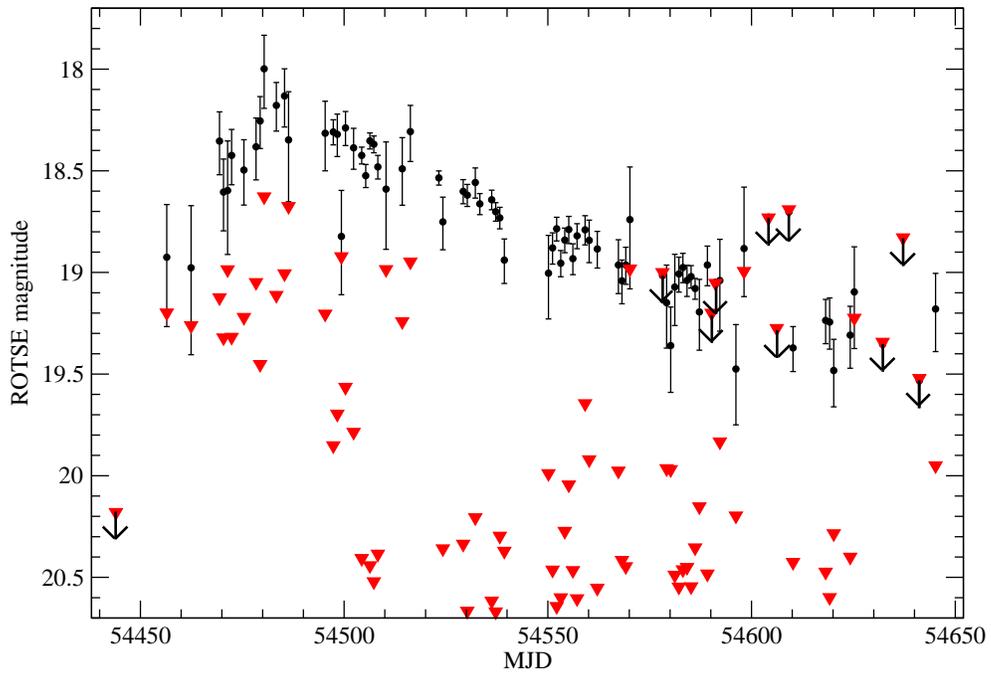}
\caption{The ROTSE unfiltered light curve of SN~2008am. The red-filled triangles represent
the sensitivity limits. Upper limits lie on their contemporaneous sensitivity points and are denoted by downward arrows.}
\end{center}
\end{figure}

\begin{figure}
\begin{center}
\includegraphics[angle=270,width=15cm]{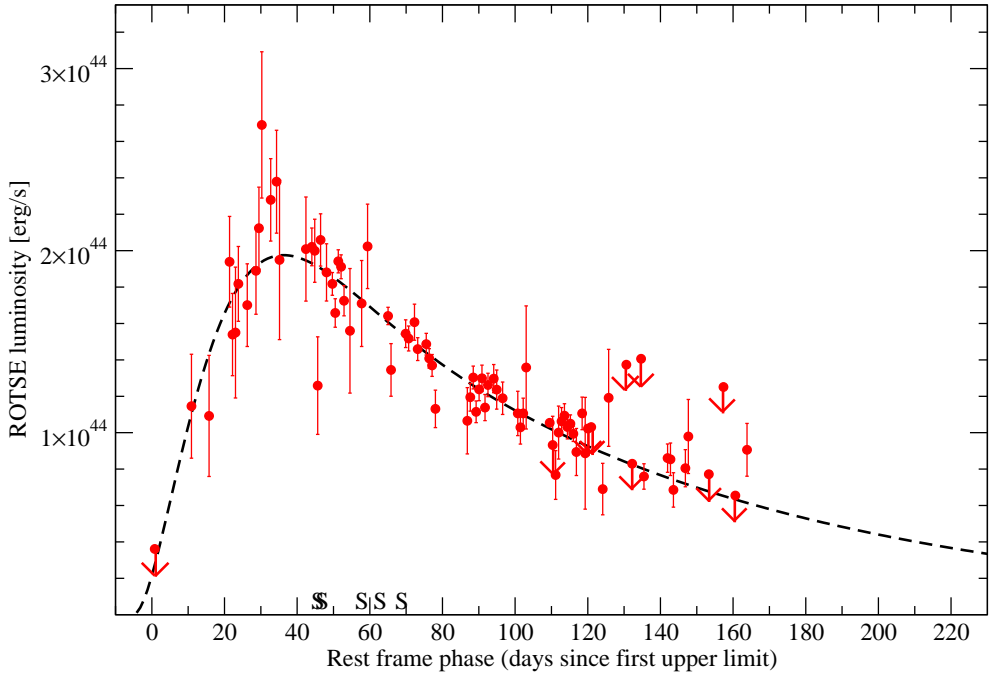}
\caption{Fit of a simple radioactive-decay diffusion model to the ROTSE light curve of SN~2008am (filled circles). The ``S"
letters above the time axis denote the spectroscopic epochs.
The best-fitting explosion date is MJD~54438.8, 5.2 days before the first upper limit, 14 days before the first
detection and ~34 days before maximum in the rest frame.
The derived nickel and ejecta mass are $M_{Ni} =$~19$M_{\odot}$ and $M_{ej} =$~0.2$M_{\odot}$ (for Thompson scattering
opacity $\kappa =$~0.4~$cm^{2}~g^{-1}$ and velocity $v_{sh} =$~1000~km~s$^{-1}$).
The model provides a means to constrain the explosion date, but is clearly not a valid physical model for the explosion (see text).}
\end{center}
\end{figure}

\begin{figure}
\begin{center}
\includegraphics[angle=0,width=15cm]{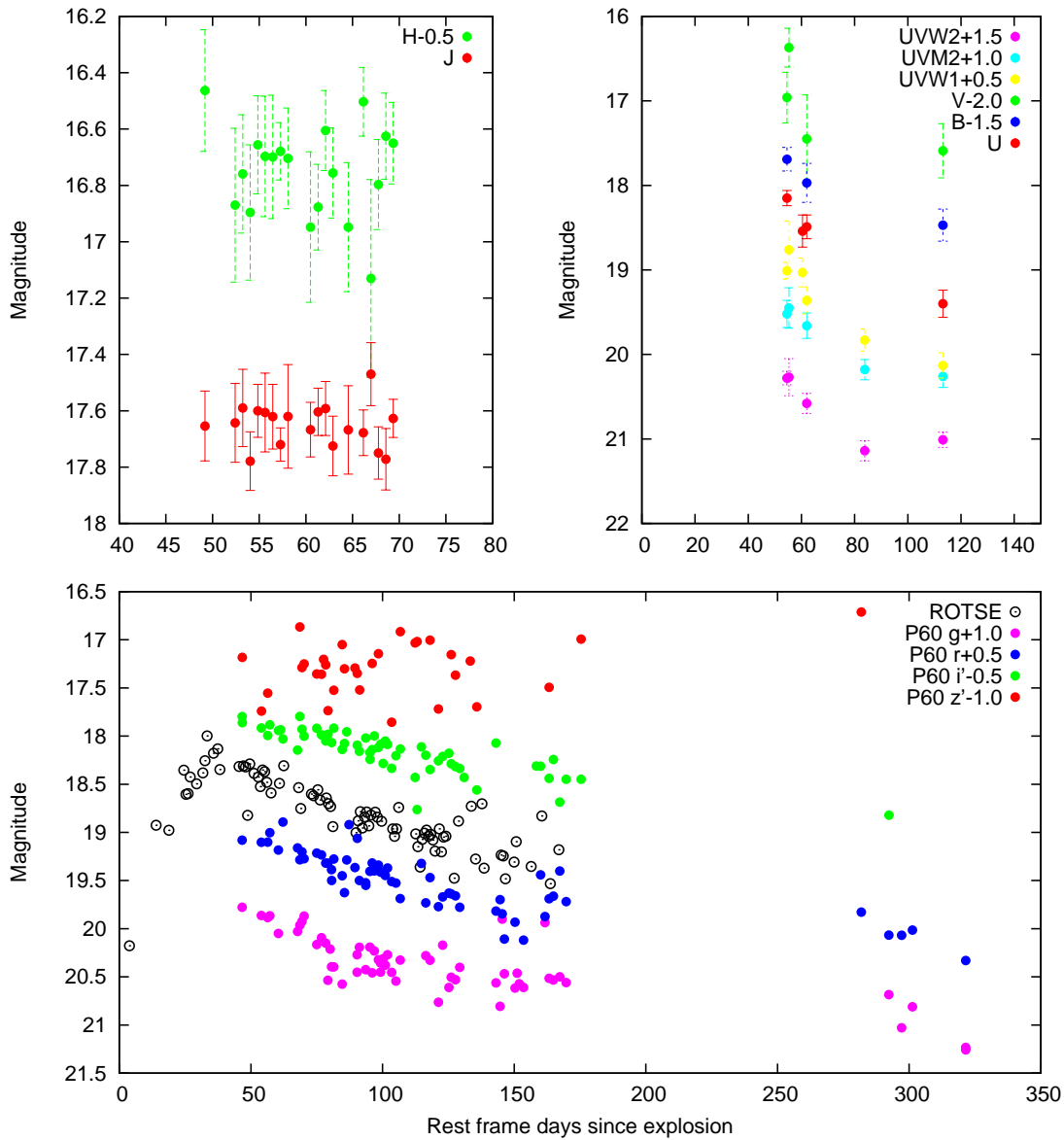}
\caption{The rest-frame light curves of SN~2008am from the IR to the UV.
{\it Top left Panel:} The PAIRITEL IR J and H band light curves. 
{\it Top right panel:} The {\it Swift} UVOT optical and UV light curves.
{\it Bottom panel:} The ROTSE unfiltered (open circles) and the optical P60 (filled circles) light curves. 
The data have been offset for clarity.}
\end{center}
\end{figure}

\begin{figure}
\begin{center}
\includegraphics[angle=270,width=15cm]{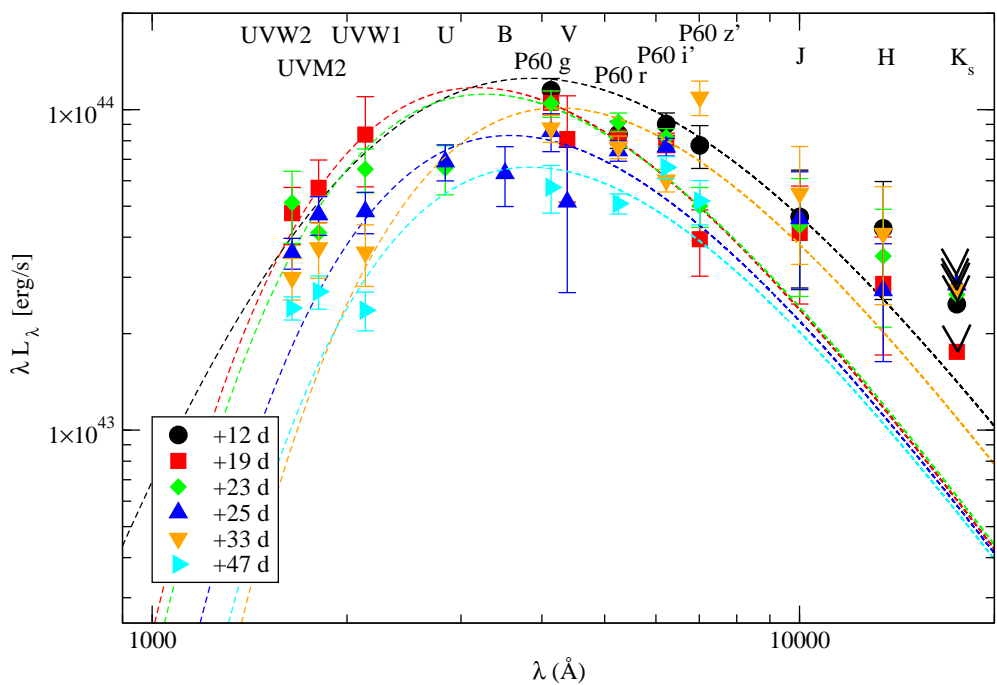}
\caption{The six (five UVOIR plus one OIR) SEDs of SN~2008am and their corresponding single temperature black-body fits. 
The black arrows in the rest-frame K-band fluxes indicate upper limits. All the phases in the inset refer to the rest frame
time in days since maximum. The photometric filters at their redshifted peak wavelength position are also indicated.}
\end{center}
\end{figure}

\begin{figure}
\begin{center}
\includegraphics[angle=270,width=15cm]{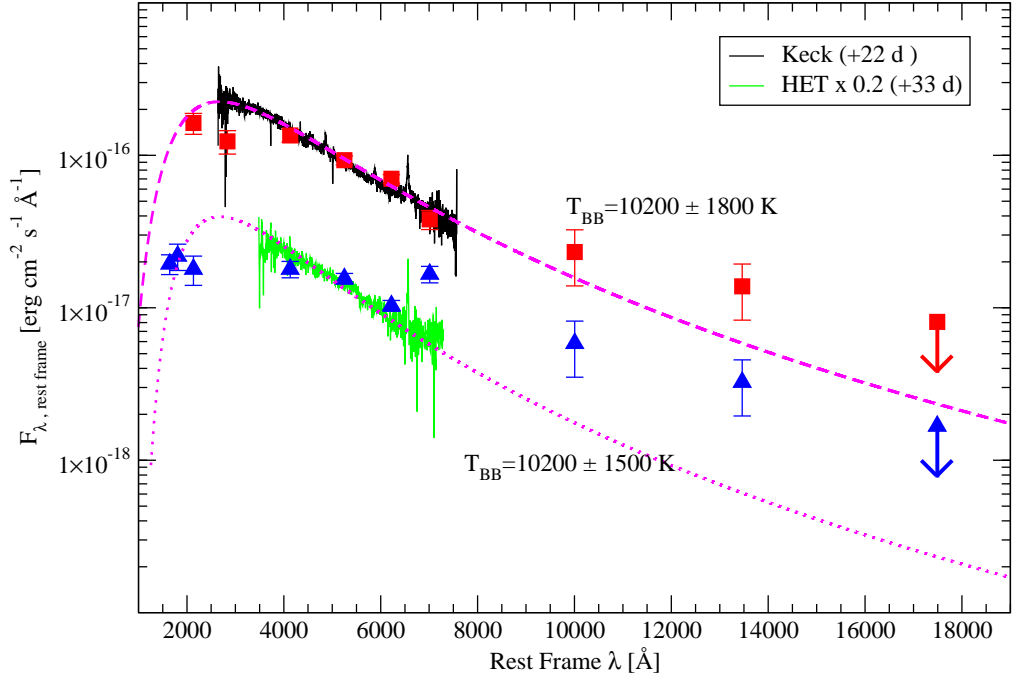}
\caption{Black-body curve fits for SN~2008am for two epochs for which there is photometric 
and spectroscopic overlap. The photometric SED for rest-frame day +23 (red squares) is constructed using UVOT+P60+PAIRITEL J data and the SED
for rest-frame day +33 (blue triangles) uses P60+PAIRITEL data. 
The best-fit black-body curves for the photometric data are given by the dashed line (+23d) and the dotted line (+33d). The corresponding
temperatures (see Table 6) are given next to the respective black-body curves. 
The temperatures of the black-body fits to the spectra are 11,100 K on day +22 and 12,200 K on day +33.
The HET LRS spectrum
has been scaled down by a factor of five for presentation.}
\end{center}
\end{figure}

\begin{figure}
\begin{center}
\includegraphics[angle=270,width=15cm]{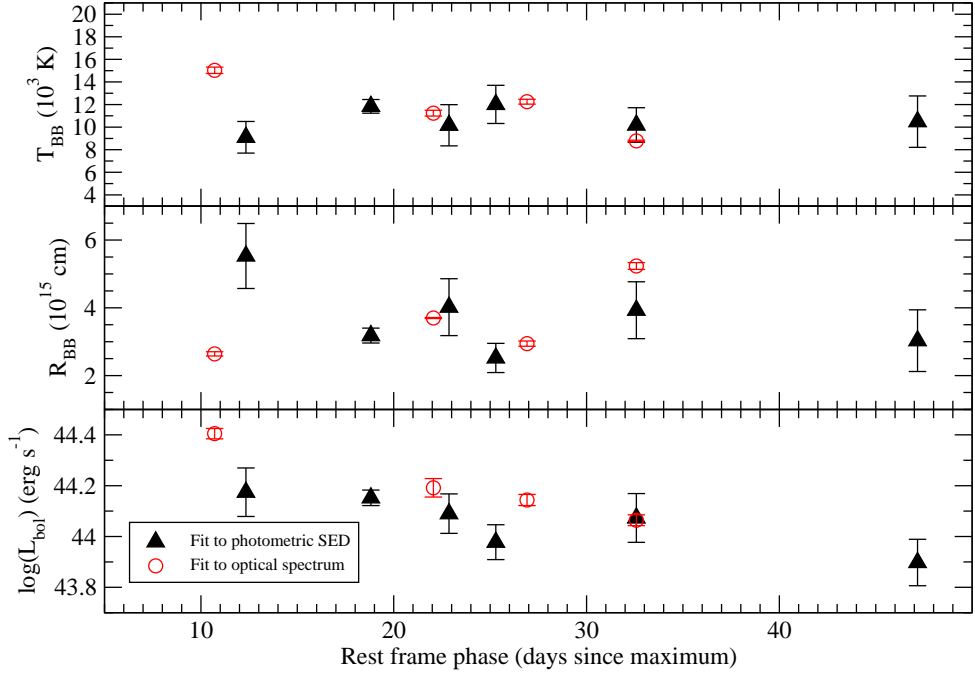}
\caption{The evolution of the effective black-body temperature $T_{bb}$ ({\it upper panel}), the effective black-body
radius $R_{bb}$ ({\it middle panel}) and the bolometric luminosity $L_{bol}$ ({\it lower panel}) of SN~2008am as estimated by 
black-body fits to the rest frame photometric SEDs for the six epochs for which we have UVOIR and OIR data.
The filled triangles refer to the fits to the photometric SEDs. The open circles 
represent black-body fits to optical spectra at similar phases.}
\end{center}
\end{figure}

\begin{figure}
\begin{center}
\includegraphics[angle=270,width=15cm]{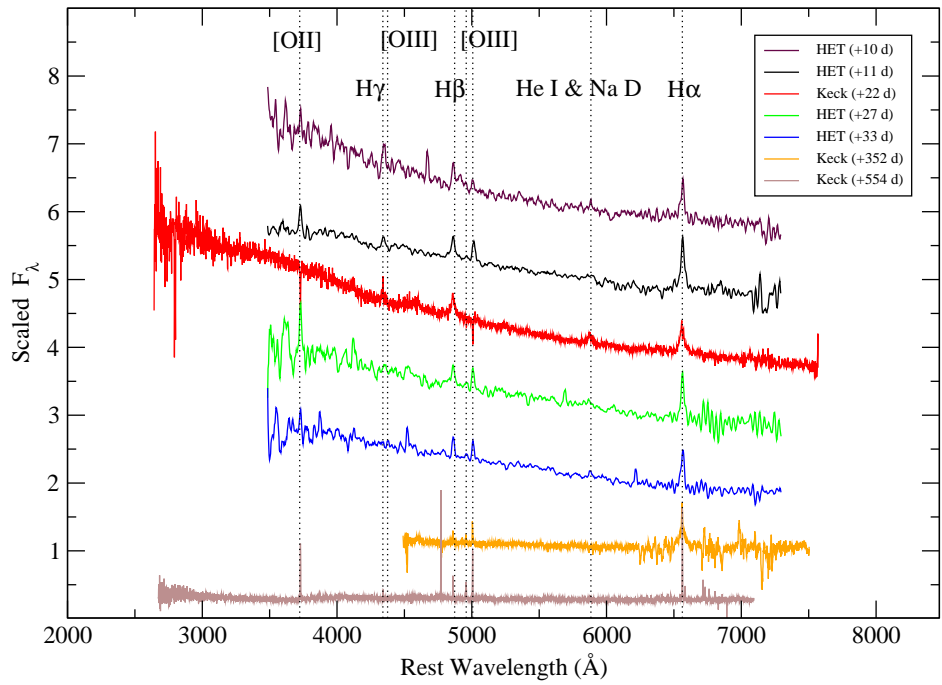}
\caption{The spectral evolution of SN~2008am in the rest frame. The phases given in the inset refer
to the rest frame time in days since maximum. The dotted vertical lines mark the positions
of the H, He and O features. Scaling and offsets to the flux values of the original spectra have been applied for clarity.}
\end{center}
\end{figure}

\begin{figure}
\begin{center}
\includegraphics[angle=0,width=12cm]{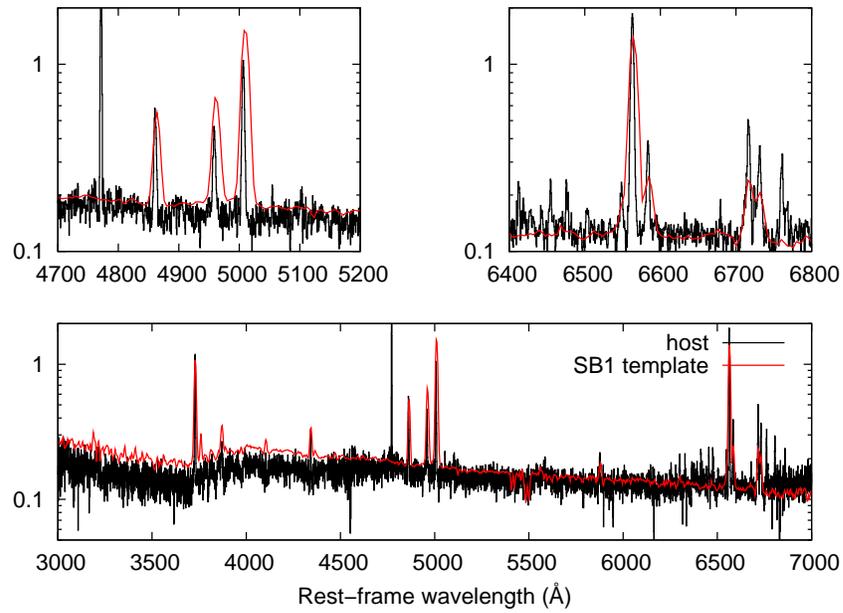}
\caption{The late +554d Keck spectrum of the host of SN~2008am (bottom panel; black solid curve) compared to an 
SB1 galaxy template (Kinney et al. 1996) (red solid curve). The H$\beta$ and [OIII] region
of the spectrum (top left). The H$\alpha$, [NII] and [SII] region of the spectrum (top right).}
\end{center}
\end{figure}

\begin{figure}
\begin{center}
\includegraphics[angle=270,width=15cm]{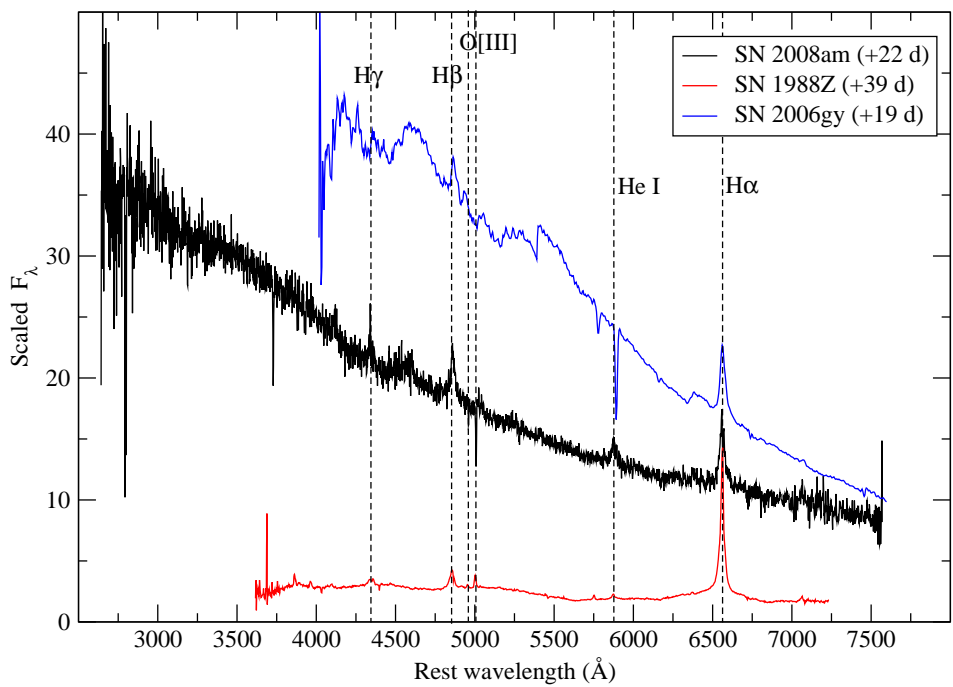}
\caption{Comparison of the Keck spectrum of SN~2008am +22d after maximum (rest frame) and the
spectrum of the classical Type IIn supernova SN~1988Z in the rest frame +39d after maximum 
(Stathakis \& Sadler 1991). The spectrum of the SLSN 2006gy at a similar rest-frame phase (+19d after maximum)
is also shown (Smith et al. 2010).
The rest frame positions of the emission lines of H, He I (5876 \AA) and [OIII] 
$\lambda \lambda$~4959, 5007 \AA ~are indicated with dashed vertical lines.}
\end{center}
\end{figure}

\begin{figure}
\begin{center}
\includegraphics[angle=270,width=15cm]{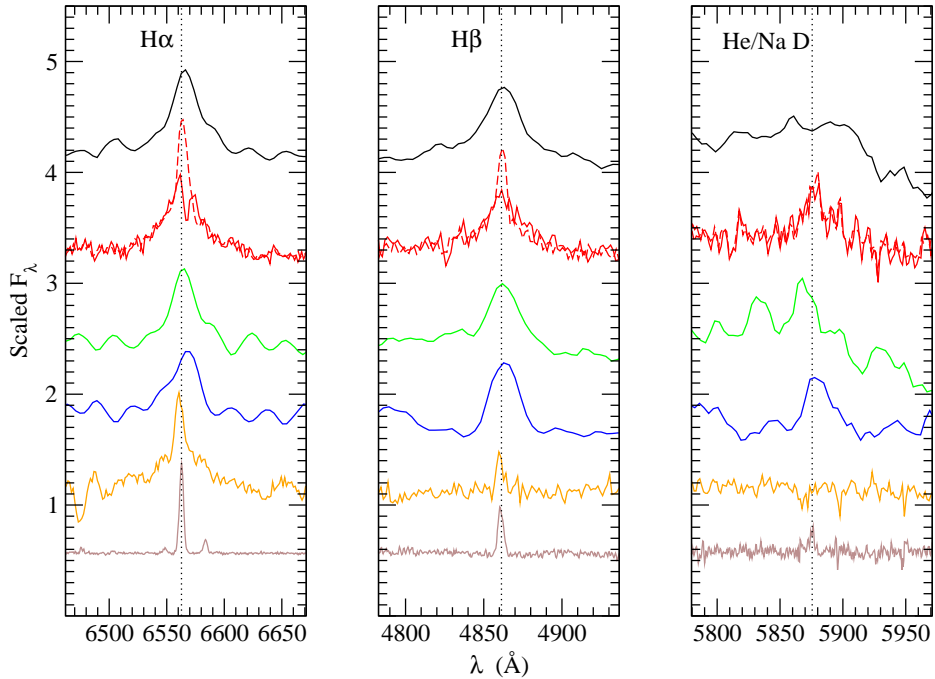}
\caption{The evolution of the H$\alpha$ (left panel), H$\beta$ (middle panel) and He/Na D (right panel) line profiles of SN~2008am. 
The five spectra correspond to +12d, +22d, +27d, +33d and +352d from rest frame maximum respectively, from top to bottom. The final spectrum
is of the host galaxy +554 rest frame days after maximum. 
The vertical dotted lines show the rest-frame location of H$\alpha$, H$\beta$, and He I $\lambda$~5876 \AA,~respectively.
For the Keck +22d spectrum we show the line profiles with (solid red curve) and without (dashed red curve) the effects of host
subtraction.}
\end{center}
\end{figure}

\begin{figure}
\begin{center}
\includegraphics[angle=0,width=12cm]{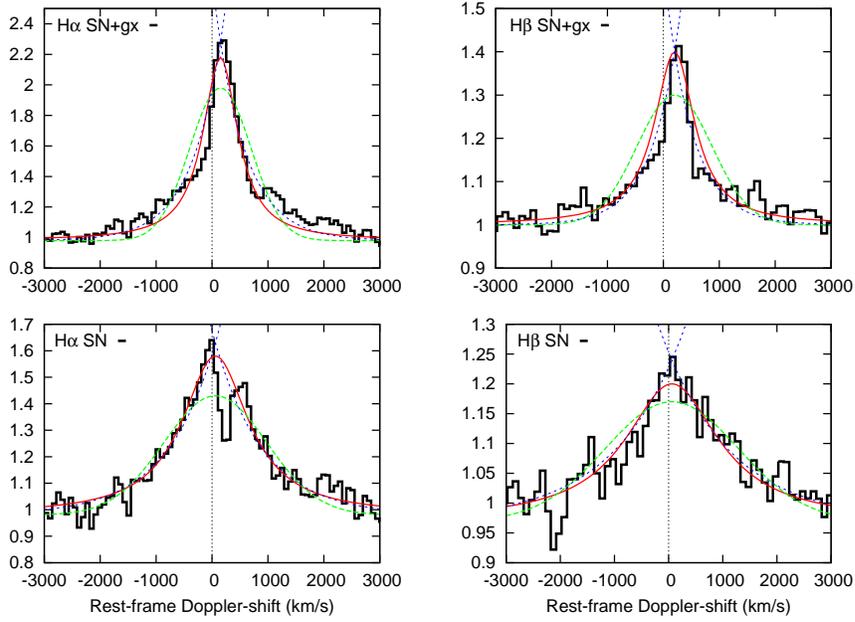}
\caption{Gaussian (dashed green curve), Lorentzian (solid red curve) and exponential (dotted blue curve) 
fits to the H$\alpha$ (upper and lower left panels) and H$\beta$ (upper and lower right panels) line profiles from the 
Keck spectrum +22d rest-frame days after maximum. In each case the upper panel shows the fits to the line profile not
corrected for host contribution and the lower panel shows the fits to the host subtracted line profile.
Lorentzian profiles provide the best overall fit to the observed line profiles (see text).}
\end{center}
\end{figure}

\begin{figure}
\begin{center}
\includegraphics[angle=270,width=15cm]{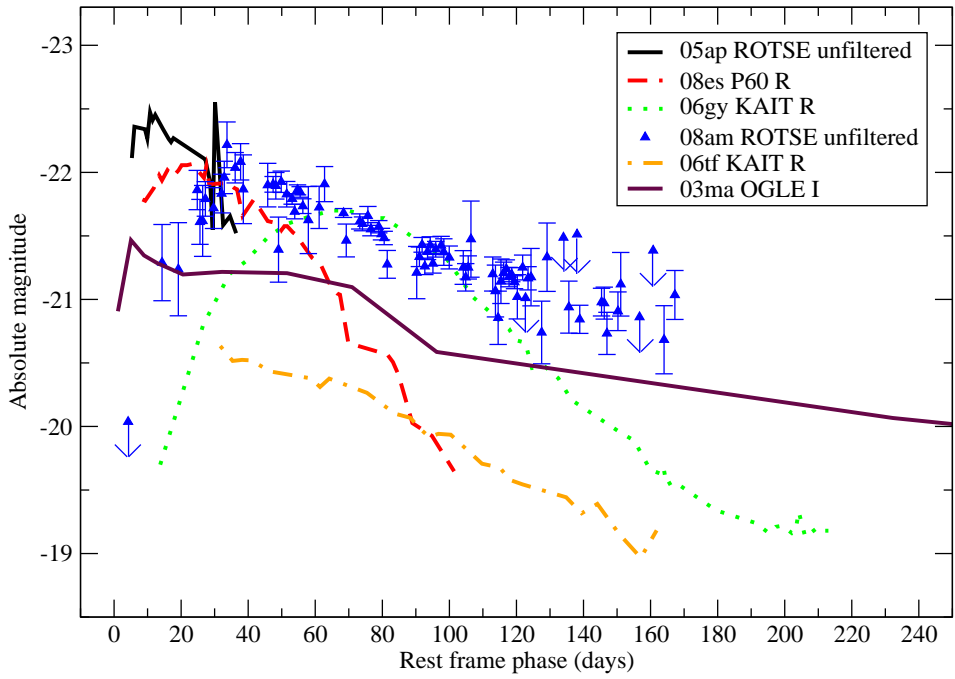}
\caption{Comparison of the rest frame light curve of SN~2008am with those of other luminous supernovae: SN~2003ma (Rest et al. 2009), 
SN~2005ap (Quimby et al. 2007a), SN~2006gy (Smith et al. 2007), 
SN~2006tf (Smith et al. 2008) and SN~2008es (Gezari et al. 2009).}
\end{center}
\end{figure}

\begin{figure}
\begin{center}
\includegraphics[angle=270,width=15cm]{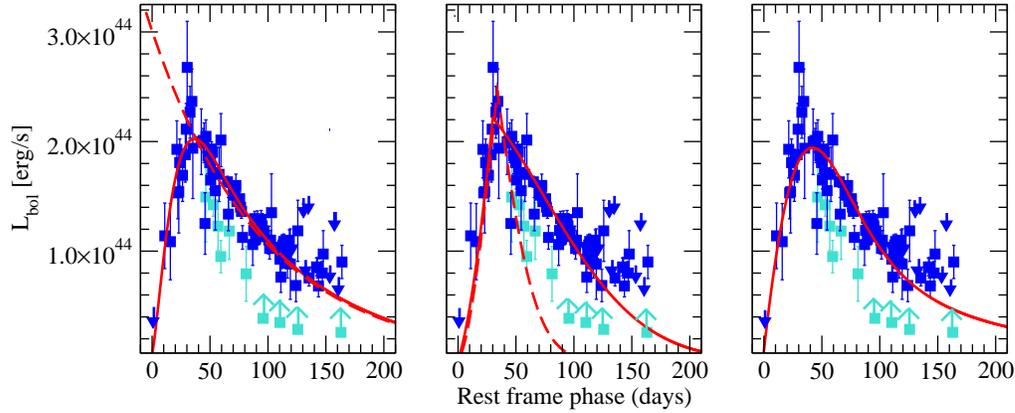}
\caption{Characteristic fits of various supernova light curve models to the ROTSE light curve of SN~2008am (filled blue squares; downwards
arrows represent upper limits).
See text (\S 4) for details on the model fitting and the derived best fitting parameters of each model.
The light blue data points correspond to the pseudo-bolometric light curve of SN~2008am derived by
the SED fits for comparison.
{\it Left panel:} Fit of a radioactive decay model (solid red curve) with $M_{Ni} =$~19~$M_{\odot}$ and $t_{d} = $~41d. The 
radioactive energy decay rate for the same amount of radioactive nickel is shown for reference (dashed red curve).
{\it Middle panel:} Fit of a shell-shock diffusion model (optically thick CSM, or hybrid model) 
for $t_{d} = t_{max} =$~34d (dashed red curve) and for $t_{d}>t_{max}$ (solid red curve). 
The $t_{d} = t_{max}$ model cannot fit the light curve of SN~2008am.
{\it Right panel:} Fit of a magnetar spin-down model (solid red curve).}
\end{center}
\end{figure}

\clearpage
\setcounter{table}{0}
\begin{deluxetable}{lcccccc}
\tabletypesize{\tiny}
\tablewidth{0pt}
\tablecaption{ROTSE-IIIb unfiltered photometry of SN2008am}
\tablehead{
\colhead {MJD$^{a}$} &
\colhead {UT date} &
\colhead {Magnitude$^{b}$} &
\colhead {Error}\\}
\startdata
54436.13&2007  Dec  2.13  &(20.17)&-  \\
54443.13&2007  Dec  9.13  &18.92&0.30  \\
54448.63&2007  Dec  14.13  &18.97&0.36  \\
54469.54&2008  Jan  4.54  &18.35&0.15  \\
54470.54&2008  Jan  5.54  &18.60&0.17  \\
54471.54&2008  Jan  6.54  &18.59&0.27  \\
54472.54&2008  Jan  7.54  &18.42&0.13  \\
54475.41&2008  Jan  10.41  &18.49&0.16 \\
54478.54&2008  Jan  13.54 &18.38&0.15 \\
54479.54&2008  Jan  14.54 &18.25&0.12 \\
54480.36&2008  Jan  15.36 &17.99&0.17  \\
54483.54&2008  Jan  18.54 &18.17&0.12 \\
54485.54&2008  Jan  20.54 &18.13&0.14 \\
54486.54&2008  Jan  21.54 &18.34&0.27 \\
54495.54&2008  Jan  30.54 &18.31&0.17 \\
54497.54&2008  Feb  1.54  &18.30&0.06 \\
54498.54&2008  Feb  2.54  &18.32&0.10  \\
54499.95&2008  Feb  3.95  &18.82&0.25 \\
54500.95&2008  Feb  4.95  &18.28&0.08 \\
54502.95&2008  Feb  6.95  &18.38&0.10 \\
54504.95&2008  Feb  8.95  &18.42&0.04 \\
54505.95&2008  Feb  9.95  &18.52&0.05 \\
54506.95&2008  Feb  10.95 &18.35&0.03 \\
54507.95&2008  Feb  11.95 &18.36&0.04 \\
54508.95&2008  Feb  12.95 &18.48&0.05 \\
54510.95&2008  Feb  14.95 &18.59&0.26	\\
54514.95&2008  Feb  18.95 &18.49&0.16 \\
54516.95&2008  Feb  20.95 &18.30&0.13 \\
54523.95&2008  Feb  27.95 &18.53&0.03 \\
54524.95&2008  Feb  28.95 &18.75&0.12 \\
54529.41&2008  Mar  4.41  &18.60&0.05  \\
54530.41&2008  Mar  5.41  &18.61&0.05 \\
54532.41&2008  Mar  7.41  &18.55&0.01 \\
54533.41&2008  Mar  8.41  &18.66&0.05 \\
54536.41&2008  Mar  11.41 &18.64&0.04 \\
54537.41&2008  Mar  12.41 &18.70&0.04 \\
54538.41&2008  Mar  13.41 &18.73&0.05  \\
54539.41&2008  Mar  14.41 &18.93&0.10  \\
54550.41&2008  Mar  25.41 &19.00&0.20  \\
54551.41&2008  Mar  26.41 &18.87&0.07  \\
54552.41&2008  Mar  27.41 &18.78&0.05 \\
54553.41&2008  Mar  28.41 &18.95&0.07 \\
54554.41&2008  Mar  29.41 &18.84&0.06 \\
54555.41&2008  Mar  30.41 &18.78&0.06 \\
54556.41&2008  Mar  31.41 &18.93&0.07 \\
54557.41&2008  Apr  1.41  &18.81&0.06 \\
54559.83&2008  Apr  3.83  &18.79&0.07  \\
54560.83&2008  Apr  4.83  &18.84&0.10  \\
54562.83&2008  Apr  6.83  &18.88&0.09 \\
54567.83&2008  Apr  11.83 &18.96&0.13 \\
54568.83&2008  Apr  12.83 &19.04&0.10  \\
54569.83&2008  Apr  13.83 &18.96&0.09  \\
54570.83&2008  Apr  14.83 &18.74&0.29  \\
54578.83&2008  Apr  22.83 &(19.01)&-  \\
54579.83&2008  Apr  23.83 &19.14&0.20	\\
54580.83&2008  Apr  24.83 &19.35&0.20	\\
54581.83&2008  Apr  25.83 &19.07&0.17  \\
54582.83&2008  Apr  26.83 &19.00&0.08 \\
54583.83&2008  Apr  27.83 &18.97&0.07 \\
54584.83&2008  Apr  28.83 &19.03&0.07  \\
54585.83&2008  Apr  29.83 &19.02&0.05	\\
54586.83&2008  Apr  30.83 &19.07&0.05 \\
54587.83&2008  May  1.83  &19.19&0.17  \\
54589.24&2008  May  3.24  &18.96&0.09 \\
54590.24&2008  May  4.24  &(19.20)&- \\
54591.24&2008  May  5.24  &(19.04)&-  \\
54592.24&2008  May  6.24  &19.03&0.22  \\
54596.24&2008  May  10.24 &19.47&0.24 \\
54598.24&2008  May  12.24 &18.88&0.26\\
54604.24&2008  May  18.24 &(18.72)&-\\
54606.24&2008  May  20.24 &(19.27)&- \\
54609.24&2008  May  23.24 &(18.70)&-  \\
54610.24&2008  May  24.24 &19.37&0.11\\
54618.24&2008  Jun  1.24  &19.23&0.10 \\
54619.65&2008  Jun  2.65  &19.24&0.12 \\
54620.65&2008  Jun  3.65  &19.48&0.16\\
54624.65&2008  Jun  7.65  &19.30&0.15\\
54625.65&2008  Jun  8.65  &19.09&0.25 \\
54632.65&2008  Jun  15.65 &(19.35)&-  \\
54637.65&2008  Jun  20.65 &(18.82)&-\\
54641.65&2008  Jun  24.65 &(19.53)&- \\
54645.65&2008  Jun  28.65 &19.17&0.19 \\
\enddata 
\tablecomments{$^{a}$ MJD values refer to the observer frame, $^{b}$ Observed value; not
corrected for extinction. The values in parentheses represent upper limits.}
\end{deluxetable}

\setcounter{table}{1}
\begin{deluxetable}{lcccccc}
\tabletypesize{\tiny}
\tablewidth{0pt}
\tablecaption{PAIRITEL IR photometry of SN2008am}
\tablehead{
\colhead{MJD$^{a}$} &
\colhead{UT date} &
\colhead{{\it J}-band$^{b}$} &
\colhead{{\it H}-band$^{b}$} &
\colhead{{\it K$_{s}$}-band$^{b}$} \\ }

\startdata
54499.95 & 2008  Feb  3.95  & 17.65(0.12) &16.96(0.22)&16.78(0.31)\\
54502.95 & 2008  Feb  6.95  & 17.64(0.14) &17.37(0.27)&16.97(0.30)\\
54503.95 & 2008  Feb  7.95  & 17.59(0.14) &17.26(0.21)&16.69(0.22)\\
54504.95 & 2008  Feb  8.95  & 17.78(0.10) &17.40(0.24)&17.15(0.35)\\
54505.95 & 2008  Feb  9.95  & 17.60(0.09) &17.16(0.17)&16.61(0.17)\\
54506.95 & 2008  Feb  10.95 & 17.61(0.14) &17.20(0.21)&17.01(0.34)\\
54507.95 & 2008  Feb  11.95 & 17.62(0.12) &17.20(0.22)&16.88(0.31)\\
54508.95 & 2008  Feb  12.95 & 17.72(0.06) &17.18(0.10)&16.71(0.21)\\
54509.95 & 2008  Feb  13.95 & 17.62(0.18) &17.20(0.18)&16.78(0.30)\\
54510.95 & 2008  Feb  14.95 & 17.67(0.10) &17.45(0.27)&16.63(0.21)\\
54513.95 & 2008  Feb  17.95 & 17.60(0.08) &17.38(0.15)&17.06(0.31)\\
54514.95 & 2008  Feb  18.95 & 17.59(0.10) &17.11(0.14)&16.70(0.20)\\
54515.95 & 2008  Feb  19.95 & 17.73(0.11) &17.26(0.16)&16.89(0.20)\\
54516.95 & 2008  Feb  20.95 & 17.67(0.16) &17.45(0.23)&16.91(0.34)\\
54520.95 & 2008  Feb  24.95 & 17.68(0.08) &17.00(0.12)&16.67(0.19)\\
54521.95 & 2008  Feb  25.95 & 17.47(0.11) &17.63(0.35)&16.68(0.25)\\
54522.95 & 2008  Feb  26.95 & 17.75(0.09) &17.30(0.16)&16.96(0.27)\\
54523.95 & 2008  Feb  27.95 & 17.77(0.11) &17.13(0.15)&16.74(0.27)\\
54524.95 & 2008  Feb  28.95 & 17.63(0.07) &17.15(0.15)&17.08(0.24)\\	      
\enddata 
\tablecomments{$^{a}$ MJD values refer to the observer frame. $^{b}$ Observed value; not
corrected for extinction. The numbers in the parentheses represent the estimated errors of the measured values.}
\end{deluxetable}

\setcounter{table}{2}
\begin{deluxetable}{lcccccc}
\tabletypesize{\tiny}
\tablewidth{0pt}
\tablecaption{Northern standard stars used for the P60 calibration. The magnitudes
have been tied to the SDSS g$^{\prime}$r$^{\prime}$i$^{\prime}$z$^{\prime}$ system.}
\tablehead{
\colhead {ID} &
\colhead {g$^{\prime}$} &
\colhead {r$^{\prime}$} &
\colhead {i$^{\prime}$} &
\colhead {z$^{\prime}$}\\}
\startdata
SDSS J122838.04+153354.4 & 17.82 & 17.05 & 16.74 & 16.60 \\
SDSS J122833.68+153505.2 & 18.18 & 17.27 & 16.95 & 16.79 \\
SDSS J122825.56+153450.2 & 17.07 & 16.15 & 15.85 & 15.69 \\
SDSS J122851.30+153427.7 & 17.62 & 17.37 & 17.27 & 17.22 \\
SDSS J122849.58+153536.8 & 16.33 & 15.95 & 15.81 & 15.75 \\      
\enddata 
\end{deluxetable}

\clearpage
\setcounter{table}{3}
\begin{deluxetable}{lcccccc}
\tabletypesize{\tiny}
\tablewidth{0pt}
\tablecaption{P60 photometry of SN2008am}
\tablehead{
\colhead {MJD$^{a}$} &
\colhead {UT date} &
\colhead {Filter} &
\colhead {Magnitude$^{b}$} &
\colhead {Error}\\}
\startdata
54496.54 &2008  Jan  31.54&g$^{\prime}$&18.78 &0.11 \\
54505.95 &2008  Feb  9.95 &g$^{\prime}$&18.86 &0.10 \\
54508.95 &2008  Feb  12.95&g$^{\prime}$&18.89 &0.11 \\
54509.95 &2008  Feb  13.95&g$^{\prime}$&18.89 &0.11 \\
54513.95 &2008  Feb  17.95&g$^{\prime}$&19.05 &0.15 \\
54522.95 &2008  Feb  26.95&g$^{\prime}$&19.03 &0.11 \\
54523.95 &2008  Feb  27.95&g$^{\prime}$&18.96 &0.10 \\
54524.95 &2008  Feb  28.95&g$^{\prime}$&18.93 &0.09 \\
54525.95 &2008  Feb  29.95&g$^{\prime}$&18.87 &0.11 \\
54531.41 &2008  Mar  6.41 &g$^{\prime}$&19.17 &0.12 \\
54533.41 &2008  Mar  8.41 &g$^{\prime}$&19.09 &0.11 \\
54535.41 &2008  Mar  10.41&g$^{\prime}$&19.15 &0.13 \\
54537.41 &2008  Mar  12.41&g$^{\prime}$&19.54 &0.14 \\
54538.41 &2008  Mar  13.41&g$^{\prime}$&19.21 &0.11 \\
54538.41 &2008  Mar  13.41&g$^{\prime}$&19.40 &0.18 \\
54539.41 &2008  Mar  14.41&g$^{\prime}$&19.40 &0.14 \\
54543.41 &2008  Mar  18.41&g$^{\prime}$&19.58 &0.30 \\
54550.41 &2008  Mar  25.41&g$^{\prime}$&19.45 &0.12 \\
54550.41 &2008  Mar  25.41&g$^{\prime}$&19.27 &0.12 \\  	   
54551.41 &2008  Mar  26.41&g$^{\prime}$&19.19 &0.12 \\ 
54554.41 &2008  Mar  29.41&g$^{\prime}$&19.43 &0.12 \\
54556.41 &2008  Mar  31.41&g$^{\prime}$&19.19 &0.11 \\
54557.83 &2008  Apr  1.83 &g$^{\prime}$&19.46 &0.12 \\
54558.83 &2008  Apr  2.83 &g$^{\prime}$&19.23 &0.09 \\
54560.83 &2008  Apr  4.83 &g$^{\prime}$&19.32 &0.12 \\
54561.83 &2008  Apr  5.83 &g$^{\prime}$&19.45 &0.11 \\
54561.83 &2008  Apr  5.83 &g$^{\prime}$&19.36 &0.13 \\
54562.83 &2008  Apr  6.83 &g$^{\prime}$&19.31 &0.10 \\
54563.83 &2008  Apr  7.83 &g$^{\prime}$&19.38 &0.12 \\
54564.83 &2008  Apr  8.83 &g$^{\prime}$&19.27 &0.15 \\
54566.83 &2008  Apr  10.83&g$^{\prime}$&19.45 &0.13 \\
54568.83 &2008  Apr  12.83&g$^{\prime}$&19.54 &0.13 \\
54570.83 &2008  Apr  14.83&g$^{\prime}$&19.33 &0.19 \\
54582.83 &2008  Apr  26.83&g$^{\prime}$&19.28 &0.19 \\
54584.83 &2008  Apr  28.83&g$^{\prime}$&19.33 &0.15 \\
54588.24 &2008  May  2.24 &g$^{\prime}$&19.76 &0.13 \\
54590.24 &2008  May  4.24 &g$^{\prime}$&19.17 &0.13 \\
54593.24 &2008  May  7.24 &g$^{\prime}$&19.61 &0.12 \\
54594.24 &2008  May  8.24 &g$^{\prime}$&19.50 &0.13 \\
54596.24 &2008  May  10.24&g$^{\prime}$&19.53 &0.14 \\
54598.24 &2008  May  12.24&g$^{\prime}$&19.40 &0.17 \\
54615.24 &2008  May  29.24&g$^{\prime}$&19.56 &0.16 \\
54617.24 &2008  May  31.24&g$^{\prime}$&19.81 &0.16 \\
54618.65 &2008  Jun  1.65 &g$^{\prime}$&18.90 &0.19 \\
54619.65 &2008  Jun  2.65 &g$^{\prime}$&19.47 &0.15 \\
54624.65 &2008  Jun  7.65 &g$^{\prime}$&19.62 &0.16 \\
54625.65 &2008  Jun  8.65 &g$^{\prime}$&19.46 &0.15 \\
54626.65 &2008  Jun  9.65 &g$^{\prime}$&19.57 &0.19 \\
54628.65 &2008  Jun  11.65&g$^{\prime}$&19.61 &0.21 \\
54638.65 &2008  Jun  21.65&g$^{\prime}$&18.94 &0.21 \\
54640.65 &2008  Jun  23.65&g$^{\prime}$&19.52 &0.13 \\
54642.65 &2008  Jun  25.65&g$^{\prime}$&19.53 &0.23 \\
54645.65 &2008  Jun  28.65&g$^{\prime}$&19.50 &0.18 \\
54648.06 &2008  Jul  1.06 &g$^{\prime}$&19.56 &0.19 \\
54799.19 &2008  Nov  27.19&g$^{\prime}$&19.68 &0.16 \\
54805.13 &2008  Dec  5.13 &g$^{\prime}$&20.03 &0.18 \\
54811.13 &2008  Dec  1.13 &g$^{\prime}$&19.81 &0.19 \\
54836.54 &2009  Jan  5.54 &g$^{\prime}$&20.23 &0.18 \\
54836.54 &2009  Jan  5.54 &g$^{\prime}$&20.26 &0.17 \\
54496.54 &2008  Jan  31.54&r$^{\prime}$&18.58 &0.10 \\
54505.95 &2008  Feb  9.95 &r$^{\prime}$&18.60 &0.05 \\
54508.95 &2008  Feb  12.95&r$^{\prime}$&18.60 &0.05 \\
54509.95 &2008  Feb  13.95&r$^{\prime}$&18.50 &0.07 \\
54513.95 &2008  Feb  17.95&r$^{\prime}$&18.68 &0.08 \\
54515.95 &2008  Feb  19.95&r$^{\prime}$&18.39 &0.14 \\
54522.95 &2008  Feb  26.95&r$^{\prime}$&18.66 &0.08 \\
54523.95 &2008  Feb  27.95&r$^{\prime}$&18.76 &0.07 \\
54524.95 &2008  Feb  28.95&r$^{\prime}$&18.70 &0.05 \\
54525.95 &2008  Feb  29.95&r$^{\prime}$&18.77 &0.05 \\
54531.41 &2008  Mar  6.41 &r$^{\prime}$&18.72 &0.06 \\
54533.41 &2008  Mar  8.41 &r$^{\prime}$&18.73 &0.08 \\
54535.41 &2008  Mar  10.41&r$^{\prime}$&18.82 &0.07 \\
54537.41 &2008  Mar  12.41&r$^{\prime}$&18.82 &0.07 \\
54538.41 &2008  Mar  13.41&r$^{\prime}$&18.89 &0.08 \\
54538.41 &2008  Mar  13.41&r$^{\prime}$&19.00 &0.07 \\
54539.41 &2008  Mar  14.41&r$^{\prime}$&18.78 &0.06 \\
54543.41 &2008  Mar  18.41&r$^{\prime}$&18.95 &0.12 \\
54544.41 &2008  Mar  19.41&r$^{\prime}$&19.13 &0.16 \\
54545.41 &2008  Mar  20.41&r$^{\prime}$&18.79 &0.21 \\
54546.41 &2008  Mar  21.41&r$^{\prime}$&18.42 &0.19 \\
54549.41 &2008  Mar  24.41&r$^{\prime}$&18.86 &0.14 \\
54550.41 &2008  Mar  25.41&r$^{\prime}$&18.56 &0.10 \\
54551.41 &2008  Mar  26.41&r$^{\prime}$&19.00 &0.06 \\
54554.41 &2008  Mar  29.41&r$^{\prime}$&19.02 &0.07 \\
54554.41 &2008  Mar  29.41&r$^{\prime}$&19.05 &0.07 \\
54556.41 &2008  Mar  31.41&r$^{\prime}$&18.90 &0.07 \\
54557.83 &2008  Apr  1.83 &r$^{\prime}$&18.82 &0.07 \\
54558.83 &2008  Apr  2.83 &r$^{\prime}$&18.90 &0.06 \\
54560.83 &2008  Apr  4.83 &r$^{\prime}$&18.84 &0.07 \\
54561.83 &2008  Apr  5.83 &r$^{\prime}$&18.91 &0.07 \\
54562.83 &2008  Apr  6.83 &r$^{\prime}$&18.92 &0.07 \\
54563.83 &2008  Apr  7.83 &r$^{\prime}$&18.94 &0.06 \\
54564.83 &2008  Apr  8.83 &r$^{\prime}$&18.87 &0.07 \\
54566.83 &2008  Apr  10.83&r$^{\prime}$&19.01 &0.06 \\
54568.83 &2008  Apr  12.83&r$^{\prime}$&19.03 &0.09 \\
54570.83 &2008  Apr  14.83&r$^{\prime}$&19.19 &0.16 \\
54580.83 &2008  Apr  24.83&r$^{\prime}$&18.82 &0.12 \\
54582.83 &2008  Apr  26.83&r$^{\prime}$&19.23 &0.12 \\
54584.83 &2008  Apr  28.83&r$^{\prime}$&18.97 &0.12 \\
54588.24 &2008  May  2.24 &r$^{\prime}$&19.27 &0.08 \\
54590.24 &2008  May  4.24 &r$^{\prime}$&19.17 &0.08 \\
54593.24 &2008  May  7.24 &r$^{\prime}$&19.13 &0.07 \\
54594.24 &2008  May  8.24 &r$^{\prime}$&19.14 &0.07 \\
54596.24 &2008  May  10.24&r$^{\prime}$&19.16 &0.12 \\
54598.24 &2008  May  12.24&r$^{\prime}$&19.28 &0.14 \\
54615.24 &2008  May  29.24&r$^{\prime}$&19.32 &0.12 \\
54617.24 &2008  May  31.24&r$^{\prime}$&19.20 &0.26 \\
54618.65 &2008  Jun  1.65 &r$^{\prime}$&19.35 &0.10 \\
54619.65 &2008  Jun  2.65 &r$^{\prime}$&19.61 &0.12 \\
54624.65 &2008  Jun  7.65 &r$^{\prime}$&19.43 &0.10 \\
54628.65 &2008  Jun  11.65&r$^{\prime}$&19.62 &0.18 \\
54636.65 &2008  Jun  19.65&r$^{\prime}$&18.94 &0.16 \\
54638.65 &2008  Jun  21.65&r$^{\prime}$&19.37 &0.19 \\
54640.65 &2008  Jun  23.65&r$^{\prime}$&19.19 &0.09 \\
54642.65 &2008  Jun  25.65&r$^{\prime}$&19.16 &0.10 \\
54645.65 &2008  Jun  28.65&r$^{\prime}$&18.90 &0.10 \\
54648.06 &2008  Jul  1.06 &r$^{\prime}$&19.22 &0.11 \\
54787.72 &2008  Nov  1.72 &r$^{\prime}$ &19.33 &0.17 \\
54800.71 &2008  Nov  3.71 &r$^{\prime}$ &19.57 &0.17 \\
54805.13 &2008  Dec  5.13 &r$^{\prime}$ &19.57 &0.11 \\
54811.13 &2008  Dec  1.13 &r$^{\prime}$ &19.51 &0.14 \\
54836.54 &2009  Jan  5.54 &r$^{\prime}$ &19.83 &0.13 \\
54496.54 &2008  Jan  31.54&i$^{\prime}$& 18.30& 0.08 \\ 
54496.54 &2008  Jan  31.54&i$^{\prime}$& 18.36& 0.07 \\
54505.95 &2008  Feb  9.95 &i$^{\prime}$& 18.42& 0.07 \\
54508.95 &2008  Feb  12.95&i$^{\prime}$& 18.49& 0.07 \\
54509.95 &2008  Feb  13.95&i$^{\prime}$& 18.38& 0.06 \\
54513.95 &2008  Feb  17.95&i$^{\prime}$& 18.44& 0.06 \\
54514.95 &2008  Feb  18.95&i$^{\prime}$& 18.44& 0.14 \\
54515.95 &2008  Feb  19.95&i$^{\prime}$& 18.53& 0.11 \\
54522.95 &2008  Feb  26.95&i$^{\prime}$& 18.65& 0.07 \\
54523.95 &2008  Feb  27.95&i$^{\prime}$& 18.30& 0.07 \\
54524.95 &2008  Feb  28.95&i$^{\prime}$& 18.43& 0.06 \\
54525.95 &2008  Feb  29.95&i$^{\prime}$& 18.50& 0.07 \\
54531.41 &2008  Mar  6.41 &i$^{\prime}$& 18.42& 0.07 \\
54533.41 &2008  Mar  8.41 &i$^{\prime}$& 18.49& 0.08 \\
54535.41 &2008  Mar  10.41&i$^{\prime}$& 18.55& 0.09 \\
54537.41 &2008  Mar  12.41&i$^{\prime}$& 18.48& 0.13 \\
54538.41 &2008  Mar  13.41&i$^{\prime}$& 18.57& 0.08 \\
54539.41 &2008  Mar  14.41&i$^{\prime}$& 18.42& 0.06 \\
54543.41 &2008  Mar  18.41&i$^{\prime}$& 18.64& 0.11 \\
54544.41 &2008  Mar  19.41&i$^{\prime}$& 18.58& 0.10 \\
54545.41 &2008  Mar  20.41&i$^{\prime}$& 18.46& 0.15 \\
54550.41 &2008  Mar  25.41&i$^{\prime}$& 18.59& 0.09 \\
54551.41 &2008  Mar  26.41&i$^{\prime}$& 18.66& 0.07 \\
54554.41 &2008  Mar  29.41&i$^{\prime}$& 18.52& 0.06 \\
54556.41 &2008  Mar  31.41&i$^{\prime}$& 18.67& 0.07 \\
54556.41 &2008  Mar  31.41&i$^{\prime}$& 18.74& 0.08 \\
54557.83 &2008  Apr  1.83 &i$^{\prime}$& 18.64& 0.08 \\
54558.83 &2008  Apr  2.83 &i$^{\prime}$& 18.50& 0.11 \\
54560.83 &2008  Apr  4.83 &i$^{\prime}$& 18.62& 0.07 \\
54561.83 &2008  Apr  5.83 &i$^{\prime}$& 18.58& 0.08 \\
54562.83 &2008  Apr  6.83 &i$^{\prime}$& 18.78& 0.08 \\
54563.83 &2008  Apr  7.83 &i$^{\prime}$& 18.55& 0.06 \\
54564.83 &2008  Apr  8.83 &i$^{\prime}$& 18.59& 0.08 \\
54566.83 &2008  Apr  10.83&i$^{\prime}$& 18.83& 0.08 \\
54568.83 &2008  Apr  12.83&i$^{\prime}$& 18.70& 0.08 \\
54570.83 &2008  Apr  14.83&i$^{\prime}$& 18.63& 0.15 \\
54577.83 &2008  Apr  21.83&i$^{\prime}$& 18.93& 0.18 \\
54578.83 &2008  Apr  22.83&i$^{\prime}$& 19.26& 0.23 \\
54580.83 &2008  Apr  24.83&i$^{\prime}$& 18.61& 0.09 \\
54582.83 &2008  Apr  26.83&i$^{\prime}$& 18.70& 0.10 \\
54584.83 &2008  Apr  28.83&i$^{\prime}$& 18.85& 0.14 \\
54588.24 &2008  May  2.24 &i$^{\prime}$& 18.76& 0.08 \\
54590.24 &2008  May  4.24 &i$^{\prime}$& 18.71& 0.08 \\
54593.24 &2008  May  7.24 &i$^{\prime}$& 18.68& 0.07 \\
54594.24 &2008  May  8.24 &i$^{\prime}$& 18.79& 0.10 \\
54596.24 &2008  May  10.24&i$^{\prime}$& 18.82& 0.13 \\
54598.24 &2008  May  12.24&i$^{\prime}$& 18.84& 0.12 \\
54600.24 &2008  May  14.24&i$^{\prime}$& 18.93& 0.16 \\
54606.24 &2008  May  20.24&i$^{\prime}$& 19.06& 0.16 \\
54615.24 &2008  May  29.24&i$^{\prime}$& 18.57& 0.10 \\
54634.65 &2008  Jun  17.65&i$^{\prime}$& 18.81& 0.13 \\
54636.65 &2008  Jun  19.65&i$^{\prime}$& 18.81& 0.11 \\
54640.65 &2008  Jun  23.65&i$^{\prime}$& 18.94& 0.09 \\
54642.65 &2008  Jun  25.65&i$^{\prime}$& 18.74& 0.08 \\
54645.65 &2008  Jun  28.65&i$^{\prime}$& 19.18& 0.14 \\
54648.06 &2008  Jul  1.06 &i$^{\prime}$& 18.95& 0.12 \\
54655.06 &2008  Jul  8.06 &i$^{\prime}$& 18.95& 0.15 \\
54800.71 &2008  Nov  3.71&i$^{\prime}$& 19.32& 0.15 \\
54496.54 &2008  Jan  31.54&z$^{\prime}$& 18.18& 0.14 \\
54505.95 &2008  Feb  9.95 &z$^{\prime}$& 18.74& 0.18 \\
54508.95 &2008  Feb  12.95&z$^{\prime}$& 18.55& 0.12 \\
54523.95 &2008  Feb  27.95&z$^{\prime}$& 17.87& 0.12 \\
54524.95 &2008  Feb  28.95&z$^{\prime}$& 18.29& 0.14 \\
54525.95 &2008  Feb  29.95&z$^{\prime}$& 18.25& 0.12 \\
54531.41 &2008  Mar  6.41 &z$^{\prime}$& 18.35& 0.22 \\
54533.41 &2008  Mar  8.41 &z$^{\prime}$& 18.36& 0.16 \\
54534.41 &2008  Mar  9.41 &z$^{\prime}$& 18.20& 0.13 \\
54536.41 &2008  Mar  11.41&z$^{\prime}$& 18.26& 0.11 \\
54537.41 &2008  Mar  12.41&z$^{\prime}$& 18.74& 0.22 \\
54539.41 &2008  Mar  14.41&z$^{\prime}$& 18.52& 0.14 \\
54543.41 &2008  Mar  18.41&z$^{\prime}$& 18.05& 0.13 \\
54544.41 &2008  Mar  19.41&z$^{\prime}$& 18.30& 0.18 \\
54549.41 &2008  Mar  24.41&z$^{\prime}$& 18.29& 0.20 \\
54550.41 &2008  Mar  25.41&z$^{\prime}$& 18.35& 0.12 \\
54551.41 &2008  Mar  26.41&z$^{\prime}$& 18.52& 0.12 \\
54557.83 &2008  Apr  1.83 &z$^{\prime}$& 18.25& 0.15 \\
54560.83 &2008  Apr  4.83 &z$^{\prime}$& 18.14& 0.14 \\
54566.83 &2008  Apr  10.83&z$^{\prime}$& 18.86& 0.23 \\
54570.83 &2008  Apr  14.83&z$^{\prime}$& 17.92& 0.32 \\
54577.83 &2008  Apr  21.83&z$^{\prime}$& 18.03& 0.17 \\
54578.83 &2008  Apr  22.83&z$^{\prime}$& 18.02& 0.22 \\
54584.83 &2008  Apr  28.83&z$^{\prime}$& 18.00& 0.16 \\
54588.24 &2008  May  2.24 &z$^{\prime}$& 18.72& 0.23 \\
54594.24 &2008  May  8.24 &z$^{\prime}$& 18.15& 0.23 \\
54596.24 &2008  May  10.24&z$^{\prime}$& 18.37& 0.17 \\
54603.24 &2008  May  17.24&z$^{\prime}$& 18.22& 0.25 \\
54606.24 &2008  May  20.24&z$^{\prime}$& 18.70& 0.22 \\
54640.65 &2008  Jun  23.65&z$^{\prime}$& 18.49& 0.13 \\
54655.06 &2008  Jul  8.06 &z$^{\prime}$& 17.99& 0.26 \\
54787.17 &2008  Nov  17.17&z$^{\prime}$& 17.71& 0.24 \\
\enddata 
\tablecomments{$^{a}$ MJD values refer to the observer frame, $^{b}$ Corrected for extinction}
\end{deluxetable}

\clearpage
\setcounter{table}{4}
\begin{deluxetable}{lcccccc}
\tabletypesize{\tiny}
\tablewidth{0pt}
\tablecaption{Swift UVOT photometry of SN2008am}
\tablehead{
\colhead {MJD$^{a}$} &
\colhead {UT date} &
\colhead {Filter} &
\colhead {Magnitude$^{b}$} &
\colhead {Error}\\}
\startdata
54502.95&2008  Feb   7.95&V    &18.96 &0.30\\
54503.95&2008  Feb   8.95&V    &18.37 &0.23\\
54511.95&2008  Feb  16.95&V    &19.45 &0.52\\
54575.83&2008  Apr  19.83&V    &19.59 &0.32\\
54502.95&2008  Feb   7.95&B    &19.19 &0.14\\
54511.95&2008  Feb  16.95&B    &19.47 &0.23\\
54575.83&2008  Apr  19.83&B    &19.97 &0.19\\
54502.95&2008  Feb   7.95&U    &18.15 &0.09\\
54509.95&2008  Feb  14.95&U    &18.54 &0.19\\
54511.95&2008  Feb  16.95&U    &18.49 &0.14\\
54575.83&2008  Apr  19.83&U    &19.40 &0.16\\
54502.95&2008  Feb   7.95&UVW1 &18.51 &0.10\\
54503.95&2008  Feb   8.95&UVW1 &18.26 &0.34\\
54509.95&2008  Feb  14.95&UVW1 &18.53 &0.17\\
54511.95&2008  Feb  16.95&UVW1 &18.86 &0.16\\
54538.41&2008  Mar  14.41&UVW1 &19.33 &0.13\\
54575.83&2008  Apr  19.83&UVW1 &19.63 &0.15\\
54502.95&2008  Feb   7.95&UVM2 &18.52 &0.16\\
54503.95&2008  Feb   8.95&UVM2 &18.45 &0.24\\	       
54511.95&2008  Feb  16.95&UVM2 &18.66 &0.15\\ 
54538.41&2008  Mar  14.41&UVM2 &19.18 &0.12\\
54575.83&2008  Apr  19.83&UVM2 &19.26 &0.13\\
54502.95&2008  Feb   7.95&UVW2 &18.78 &0.08\\
54503.95&2008  Feb   8.95&UVW2 &18.77 &0.22\\
54511.95&2008  Feb  16.95&UVW2 &19.08 &0.12\\
54538.41&2008  Mar  14.41&UVW2 &19.64 &0.12\\
54575.83&2008  Apr  19.83&UVW2 &19.51 &0.09\\
\enddata 
\tablecomments{$^{a}$ MJD values refer to the observer frame, $^{b}$ Observed value; not
corrected for extinction}
\end{deluxetable}

\clearpage
\setcounter{table}{5}
\begin{deluxetable}{llllllllllccccccc}
\tabletypesize{\tiny}
\tablewidth{0pt}
\tablecaption{Characteristics of the black-body fits to the photometric SEDs of SN~2008am}
\tablehead{
\colhead {Epoch (MJD)} &
\colhead {$t_{rf}$ (days)} &
\colhead {$\chi^{2}$/dof} &
\colhead {$T_{bb}$ ($10^{4}$ K)} &
\colhead {$R_{bb}$ ($10^{15}$~cm)} &
\colhead {$L_{bol}$ ($10^{44}$ erg~s$^{-1}$)} \\
}
\startdata
54495 & +12 &2.4 &  0.910  (0.140)  & 5.530 (0.960) & 1.490 (0.410)\\
54504 & +19 &1.3 &  1.200  (0.060)  & 3.190 (0.220) & 1.420 (0.100)\\
54509 & +23 &2.7 &  1.020  (0.180)  & 4.020 (0.840) & 1.230 (0.220)\\
54512 & +25 &2.4 &  1.200  (0.170)  & 2.520 (0.430) & 0.950 (0.150)\\
54521 & +33 &2.9 &  1.020  (0.150)  & 3.930 (0.840) & 1.180 (0.260)\\
54539 & +47 &3.7 &  1.050  (0.230)  & 3.030 (0.910) & 0.790 (0.250)\\ 
\enddata 
\tablecomments{The numbers in parentheses represent the estimated error of each parameter. 
The $t_{rf}$ column refers to rest-frame days since maximum.}
\end{deluxetable}

\clearpage
\setcounter{table}{6}
\begin{deluxetable}{lccccccccccc}
\tabletypesize{\tiny\tiny}
\tablewidth{0pt}
\tablecaption{Summary of the H$\alpha$ and H$\beta$ properties of SN2008am.}
\tablehead{
\colhead{Instrument$^{b}$} &
\colhead{UT date} &
\colhead{Day$^{c}$} &
\colhead{EW(H$\alpha$)} &
\colhead{FWHM(H$\alpha$)} &
\colhead{$\Delta \lambda_{0}$(H$\alpha$)} &
\colhead{F(H$\alpha$)} &
\colhead{EW(H$\beta$)} &
\colhead{FWHM(H$\beta$)} &
\colhead{$\Delta \lambda_{0}$(H$\beta$)} &
\colhead{F(H$\beta$)} \\ }
\startdata
HET-LRS   & 2008 Jan 30.1  & +11  & -44   & 25 & 150 & 3.32 & -12   & 32 & 80 & 1.40  \\
Keck-LRIS & 2008 Feb 12.0  & +22  & -27   & 39 & -4  & 1.67 & -11   & 41 & 102 & 1.07 \\
Keck-LRIS$^{d}$ & 2008 Feb 12.0  & +22  & -32   & 17 & 3  & 1.98 & -10   & 14 & 179 & 1.10 \\
HET-LRS   & 2008 Feb 18.3  & +27  & -48   & 22 & 85  & 2.55 & -11   & 27 & 67 & 1.17  \\
HET-LRS   & 2008 Feb 25.3  & +33  & -42   & 24 & 185 & 2.16 & -7    & 18 & 142 & 0.72 \\
Keck-LRIS & 2009 Mar 31.0  & +352 & -307  & 25 & -60 & 1.05 & -3    &  3 & -72 & 0.06 \\
\enddata 
\tablecomments{$^{a}$ The measured equivalent widths
and fluxes refer to the rest frame, galaxy-subtracted and de-reddened spectra. The equivalent
widths and FWHM are measured in \AA ~, the shifts of the line centers with respect to their rest
frame positions ($\Delta \lambda_{0}$) in km~s$^{-1}$ and fluxes in 10$^{-15}$ erg~cm$^{-2}$~$s^{-1}$ computed
by fitting Lorentzian profiles.
$^{b}$ The HET-LRS and Keck-LRIS wavelength are 4020-10,200 \AA and 3500-8600 \AA ~respectively. 
$^{c}$ All the values refer to the rest frame days after maximum.
$^{d}$ These estimates correspond to line profiles not corrected for host extinction.}
\end{deluxetable}

\clearpage
\setcounter{table}{7}
\begin{deluxetable}{lccccccccc}
\tabletypesize{\tiny\tiny}
\tablewidth{0pt}
\tablecaption{Gaussian, Lorentzian and exponential fits to the observed H$\alpha$ and H$\beta$ line profiles
of SN~2008am.}
\tablehead{
&&&&\colhead {H$\alpha$}&&&\colhead {H$\beta$}&&\\
\hline
\colhead{Instrument$^{a}$} &
\colhead{UT date} &
\colhead{Day$^{b}$} &
\colhead{$FWHM_{G}^{c}$} &
\colhead{$FWHM_{L}^{c}$} &
\colhead{$\sigma_{exp}^{c}$} &
\colhead{$FWHM_{G}^{c}$} &
\colhead{$FWHM_{L}^{c}$} &
\colhead{$\sigma_{exp}^{c}$} \\ }
\startdata
HET-LRS   & 2008 Jan 30.1  &  +11  &  1562 &  1143 &   883 & 1956 &  1983 &    883\\
Keck-LRIS & 2008 Feb 12.0  &  +22  &  1880 &  1441 &  1438 & 2179 &  2518 &   1283\\
Keck-LRIS$^{d}$ & 2008 Feb 12.0  &  +22  &  1073 &  755 &  551 & 1315 &  872 &   503\\
HET-LRS   & 2008 Feb 18.3  &  +27  &  1317 &  1003 &   796 & 1655 &  1657 &    796\\
HET-LRS   & 2008 Feb 25.3  &  +33  &  1351 &  1114 &   901 & 1042 &  1087 &    901\\
Keck-LRIS & 2009 Mar 31.0  &  +352 &  1780 &  1142 &   918 & 280  &  190  &    148\\
\enddata 
\tablecomments{$^{a}$ The HET-LRS and Keck-LRIS wavelength ranges are 4020-10,200 \AA~ and 3500-8600 \AA ~respectively. 
$^{b}$ All the values refer to the rest frame days after maximum. $^{c}$ The quantity $FWHM_{G}$ corresponds to the fitted Gaussian,
$FWHM_{L}$ to the fitted Lorentzian and $\sigma_{exp}$ to the fitted exponential profiles, all expressed in km~s$^{-1}$.
$^{d}$ These estimates correspond to line profiles not corrected for host extinction.}
\end{deluxetable}

\end{document}